\documentclass{sigchi}

\usepackage{balance}      
\usepackage{graphicx}     
\usepackage{ucs}		
\usepackage[utf8]{inputenc}
\usepackage[T1]{fontenc}
\usepackage{txfonts}
\usepackage{mathptmx}
\usepackage{color}
\usepackage{booktabs}
\usepackage{textcomp}
\usepackage{enumitem}
\usepackage{psfrag}
\usepackage{multirow}
\usepackage{array}
\setenumerate{noitemsep,topsep=0pt,parsep=0pt,partopsep=0pt}

\usepackage{dashrule}
\usepackage{microtype}       
\usepackage{ccicons}         

\usepackage{todonotes}

\definecolor{darkpastelgreen}{rgb}{0.01, 0.75, 0.24}
\definecolor{orange-red}{rgb}{1.0, 0.27, 0.0}

\DeclareMathAlphabet{\mathcal}{OMS}{cmsy}{m}{n}

\newcommand{\R}{\mathbb{R}}
\newcommand{\N}{\mathbb{N}}

\usepackage{soulutf8}
\usepackage{nicefrac}
\usepackage[normalem]{ulem}
\usepackage[caption=false]{subfig}
\usepackage[pdflang={en-US}]{hyperref}

\def\plaintitle{An Optimal Control Model of Mouse Pointing Using the LQR}

\def\emptyauthor{}
\def\plainkeywords{Pointing; Aimed Movements; Fitts' Law; Control Theory; LQR; Modeling; Second-order Lag; Minimum Jerk}

\makeatletter
\def\url@leostyle{
  \@ifundefined{selectfont}{
    \def\UrlFont{\sf}
  }{
    \def\UrlFont{\small\bf\ttfamily}
  }}
\makeatother
\def\pprw{8.5in}
\def\pprh{11in}

\definecolor{linkColor}{RGB}{6,125,233}
\hypersetup{
  pdftitle={\plaintitle},
  pdfauthor={\emptyauthor},
  pdfkeywords={\plainkeywords},
  pdfdisplaydoctitle=true,
  bookmarksnumbered,
  pdfstartview={FitH},
  colorlinks,
  citecolor=black,
  filecolor=black,
  linkcolor=black,
  urlcolor=linkColor,
  breaklinks=true,
  hypertexnames=false
}

\usetikzlibrary{shapes,arrows,shapes.geometric}
\tikzset{
	block/.style    = {draw, thick, rectangle},
	sum/.style      = {draw, circle},
	input/.style    = {coordinate},
	output/.style   = {coordinate}
}
\tikzset{gain/.style={draw=black,text=black,inner sep=2pt, shape = isosceles triangle,shape border rotate=180},
	split/.style={draw, circle, fill=black,inner sep=0.4mm}}

\newcommand{\inte}{$\int$}

\begin{document}

\title{\plaintitle}

\numberofauthors{1}
\author{
	\alignauthor{Florian Fischer, Arthur Fleig, Markus Klar, Lars Gr\"une, J\"org M\"uller}\\
		\affaddr{University of Bayreuth, Germany}\\
	}

\maketitle

\begin{abstract}
	In this paper we explore the Linear-Quadratic Regulator (LQR) to model movement of the mouse pointer. 
	We propose a model in which users are assumed to behave optimally with respect to a certain cost function. 
	Users try to minimize the distance of the mouse pointer to the target smoothly and with minimal effort, by simultaneously minimizing the jerk of the movement.
	We identify parameters of our model from a dataset of reciprocal pointing with the mouse.
	We compare our model to the classical minimum-jerk and second-order lag models on data from 12 users with a total of 7702 movements. 
	Our results show that our approach explains the data significantly better than either of these previous models. 
\end{abstract}




\section{Introduction}
Interaction with computers is almost always achieved through movement of the user, measured via input devices.
In the field of human motor control, there has been tremendous progress in the understanding of human movement since the 1950's and 60's, when Fitts' law \cite{Fit54,FitPet64} was published.
Arguably the most important modern theory of human motor control is optimal feedback control (OFC)~\cite{todorov02,DIEDRICHSEN201031}.
Its main strengths are \emph{versatility} (applicable to many movement tasks) and the ability to \emph{predict the entire movement} (including position, velocity, and acceleration of the end-effector over time, not just movement time) without relying on Machine Learning techniques, thus retaining \emph{comprehensibility}.
Despite its advantages, OFC models are not very well known in the field of Human-Computer Interaction (HCI), yet. 
The objective of this paper is to introduce optimal feedback control to HCI. 

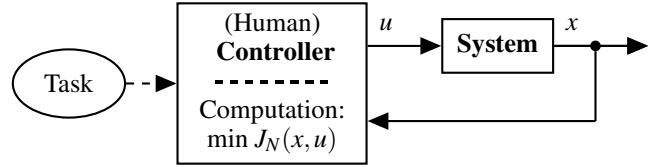
\begin{figure}[!t]
	\centering
\begin{tikzpicture}[auto, thick, node distance=1.5cm, >=triangle 45,inner sep=2mm]
\draw
node [ellipse, draw] (input1) {Task} 
node [block, right of=input1,node distance=2.7cm,inner sep=1mm] (inte1) {\begin{tabular}{c} (Human) \\ \textbf{ Controller} \\\hdashrule[0.5ex]{1.8cm}{1pt}{1mm}\\ Computation:\\ min $J_N(x,u)$ \end{tabular}}
node [block, right of=inte1,node distance=3cm,yshift=0.5 cm] (inte2) {\textbf{System}}
node [split, right of=inte2,node distance = 1.3cm] (split2) {}
node [output,right of=split2,node distance = 0.7cm] (output1) {}
node [input,below of=split2, node distance=1cm] (plhold) {};
\draw[dashed,->](input1) --  node[near start] {} (inte1);
\draw[->] ([yshift=0.5 cm]inte1.east)-- node[near start] {$u$} (inte2.west) ;
\draw[->](inte2) -- node {} (output1);
\draw[-](inte2) -- node {$x$} (split2);
\draw[->](split2) -- node{} (output1);
\draw[-](split2) -- node{} (plhold);
\draw[->](plhold) -- node{} ([yshift=-0.5 cm]inte1.east);
\end{tikzpicture}
	\caption{In our model, the user is assumed to \emph{control} the state~$x$ of the interactive system (e.g., the mouse pointer position and velocity). We assume that the user computes the control~$u$ through \emph{optimization}, i.e., by minimizing a cost function~$J_N$. In this calculation the current state is taken into account through \emph{feedback}.}
	\label{fig:genmodel}
\end{figure}

OFC is a family of computational models of (human) movement. 
These models assume that people behave rationally, i.e., optimally with respect to some cost function.
In addition, people observe the state of the environment and adjust their movement in order to accomplish a given task, in a feedback manner.
The interplay of the three main constituents of OFC, i.e., \emph{optimality}, \emph{feedback}, and \emph{control}, is displayed in Figure~\ref{fig:genmodel}.

As the figure suggests, the OFC framework is very versatile: 
Various movements such as hand or eye movements or balancing, can be explained by adjusting the \emph{System} block (and the \emph{Controller} block, if necessary). 
Various instructions, such as emphasizing speed vs.\ comfort, can be incorporated by adapting the cost function. 
Due to their feedback structure (also called \emph{closed-loop}), OFC models provide intuitive insight in how humans react to disturbances during the movement, changing targets, 
etc.

Through OFC, we aim at connecting the field of HCI better with recent advances in neighboring scientific disciplines, such as the study of human movement in motor control~\cite{SchLee05, Flash85} and neuroscience~\cite{shadmehr:2005}.

From a scientific perspective, this would strengthen the field of HCI through a deeper insight into the basic constituents of interaction.
We start from one of the simplest and most ubiquitous ways we interact with Personal Computers: pointing with a mouse. 
However, as stated above, OFC could provide a unifying framework for understanding movement in many different interactive tasks, including pointing, steering, tracking of moving targets, scrolling and zooming, with PCs, mobile devices, in AR/VR, etc.

From an engineering perspective, OFC would enable a deeper understanding of the impact of interface design parameters on the process of interaction.
In the long term, these models could be used for automated optimization of the parameters of interaction techniques.
Models of the dynamics of interaction would help in the design of input devices, from mice to VR controllers. 
Models that work in real-time could be used in predictive interfaces, which anticipate what the user wants to do and respond accordingly, such as pointing target prediction~\cite{asano2005predictive}.

To achieve our goals, we start from a well-known model from OFC theory, presented by Todorov~\cite{Todorov:Thesis}.
We believe that the best way to introduce modern motor control theory to HCI is to provide a simple model that is adapted to the above mentioned HCI purposes. 
Thus, we make several model simplifications, which we discuss below.
These allow us to use the so-called Linear-Quadratic Regulator (LQR) as the \emph{Controller} in Figure~\ref{fig:genmodel}, to calculate the optimal feedback control law.
We explore cost functions that combine the objectives of minimizing jerk, which is the derivative of acceleration, and minimizing the distance to the target.
We identify parameters of these cost functions and the underlying pointer dynamics from a dataset of reciprocal pointing~\cite{mueller17}.
We compare the ability of our model to replicate pointer movement to two other models based on the {\em second-order lag}~\cite{CroGoo83, langolf} and {\em jerk minimization}~\cite{Flash85}.
Both are suitable comparison candidates: the former model has been evaluated with the same dataset~\cite{mueller17}; the latter is an established model in motor control, which has been applied in HCI context~\cite{quinn2016}.
We compare the models on data from 12 users, with 7702 movements overall. 

Our results show that our model is able to fit the data significantly better than the other two models. 
Compared to the former, our approach can generate more symmetric and plausible velocity and acceleration profiles.
Compared to the latter, our approach allows to simultaneously model the movement well and reach the target.
Our model can predict the entire movement with only three, intuitively interpretable parameters.

\section{Related Work}
In HCI, movement, e.g., of the mouse pointer, is often reduced to summary statistics such as movement time. 
The dependency of movement time $MT$ from distance $D$ and width $W$ of targets is usually described by Fitts' law \cite{Fit54,FitPet64} as \mbox{$MT=a+b\operatorname{ID}$} with Index of Difficulty (ID) defined as \mbox{$\operatorname{ID}=\log_2(D/W+1)$} \cite{Mackenzie}, although alternatives such as Meyer's law exist \cite{meyer1988optimality}.
In HCI, Fitts' law is usually interpreted from an information theoretic perspective.
A very good explanation of this interpretation of Fitts' law has been provided by Gori et al. \cite{Gori18}. 

The kinematics and dynamics of movement are studied more rarely in HCI. 
However, in the studies of human motor control, various models describing kinematics and dynamics of human movement have been developed.

Feedback control models (also called \textit{closed-loop models}) of movement assume that people monitor and adjust their motion on a moment-to-moment basis.
These models are able to explain how users repeatedly correct errors and handle disturbances. 
An early closed-loop model (without optimization) has been provided by Crossman and Goodeve~\cite{CroGoo83}. 
They assume that users observe hand and target and adjust their velocity as a linear function of the distance, as a first-order lag. 

A simple, physically more plausible extension of the first-order lag is the second-order lag~\cite{CroGoo83,langolf}.
These dynamics can be interpreted as a spring-mass-damper system similar to that implied by the equilibrium-point theory of motor control~\cite{SchLee05}. 
A constant force is applied to the mass, such that the system moves to and remains at the target equilibrium.
This is one of the comparison models; hence, we call this approach~\emph{2OL-Eq}. 
Other models of human movement include VITE~\cite{Bullock} and the models of Plamondon \cite{plamondon1997}.

A fundamentally different approach to using such fixed-control models is to assume that humans try to behave optimally, according to a certain internalized cost function.
Flash and Hogan~\cite{Flash85} propose that humans aim to generate smooth movements by minimizing the jerk of the end effector.
We call this model \emph{MinJerk} in the following.
Although the hypothesis that people aim to minimize jerk has been questioned, see, e.g.,  Harris and Wolpert~\cite{Harris1998}, it is an established model and has been successfully used by Quinn and Zhai~\cite{quinn2016} to model the shape of gestures on a word-gesture keyboard.
The minimum-jerk model predicts a scale-invariant trajectory (as a 5th-degree polynomial), if the exact position and time of beginning and end of the movement are known. 
It can be interpreted as a trajectory planning step~\cite{todorov02} and is thus particularly appropriate for modeling movements that do not involve so-called \textit{corrective submovements}.
These have first been proposed by Woodsworth~\cite{Woo1899,Elliot2001} and typically occur after the first large movement, also called the ``surge'', towards the target~\cite{meyer1988optimality}.
Hence, while applicable for gestures, it remains to be seen whether this model can replicate mouse pointer data accurately. 
Moreover, it does not explain how people execute that trajectory, or if and how they react to disturbances, such as muscle fatigue, external perturbations, changes of the target, etc.

The theory of OFC allows to resolve the separation between trajectory planning and execution. 
Excellent overviews of recent progress in OFC theory are provided by Crevecoeur et al.\ \cite{crevecoeur} and Diedrichsen~\cite{DIEDRICHSEN201031}.
An early approach that models perturbed reach and grasp movements by using the minimum-jerk trajectory on a moment-to-moment basis was presented by Hoff and Arbib~\cite{HoffArbib93}.
A more general, more recent and better known OFC model is proposed by Todorov and Jordan~\cite{todorov02}. 
This non-deterministic model is based on an extension of the Linear-Quadratic-Gaussian Regulator (E-LQG)~\cite{Todorov:Thesis}.
It assumes that users try to reach a target at a certain time while minimizing jerk.
The biomechanical apparatus is modeled by second-order lag dynamics. 
In via-point tasks, this model qualitatively replicates movement segmentation, eye-hand coordination, visual perturbations, and other characteristics of human movement. 
A discussion about how this model, including state- and control-dependent noise, can be extended to more general reaching movements can be found in~\cite{Todorov2005StochasticOC}.

A fundamental limitation of the E-LQG model (and many other optimal control models, e.g., \cite{Flash85,Uno1989,Harris1998}) is that the exact movement time needs to be known in advance.
One way to circumvent this issue is to use infinite-horizon OFC~\cite{Jiang2011, Qian2013, Li2018}, i.e., to formulate the optimal control problem on an infinite time horizon.
In these references, this approach, in conjunction with a cost function that includes (quadratic) distance and effort costs, was used to model end-effector movement towards a target. 
The movement time then emerges from the optimal control problem.

Another strand of literature that specifically deals with the duration of movement has produced the \emph{Cost of Time} theory \cite{Hoff1994,shadmehr2010,Berret2016}.
This theory assumes that humans value time with a certain (e.g., hyperbolic or sigmoidal) cost function.
Thus, movement time is explicitly included in the cost function.

In summary, the fundamental question of human movement coordination has produced a substantial literature and deep understanding regarding the nature of human movement.
Given that almost all interaction of humans with computers involves movement, it is surprising that this knowledge is little known in HCI.
It is important to bear in mind, however, that the purposes of these models are very different from HCI.
They intend to model movement of the human body per se.
In contrast, in HCI we are less interested in how the body moves, and more interested in how virtual objects in the computer, such as mouse pointers, move.
Movement in HCI is mediated by input devices, operating systems, and programs, requires high precision, and is often learnt very well. 
Therefore, these models need to be adapted and validated regarding their ability to model movement of virtual objects such as mouse pointers in interaction. 

In the field of HCI, there are few publications with control models of mouse pointer movement.
M\"uller et al.\ \cite{mueller17} compare three feedback control models (without optimization) regarding their ability to model mouse pointer movements. 
Ziebart et al.\ \cite{Ziebart} explore the use of optimal control models for pointing target prediction.
They do not make particular a priori assumptions about the structure of the cost function.
Instead, they use a machine learning approach to fit a generic function with a large number of parameters (36) to a dataset of mouse pointer movements.
While suitable for their purposes, we are interested in gaining more insight into the structure of the cost function.
Furthermore, we believe that reducing the number of parameters (to three in our main model) reduces the risk of overfitting.

\section{Model Simplifications}
Our approach to introducing OFC theory to HCI is by providing a model that is applicable to HCI, easy enough to understand, while still showing the benefits and strengths of OFC theory.
To this end, we start with a simple model for mouse pointer movements that we validate on an HCI dataset.
Based on this initial introduction of OFC to HCI, in the future we plan to incorporate extensions proposed in the motor control literature, such as sensorimotor noise and Cost of Time theory.

Our model is inspired by Todorov's E-LQG model~\cite{Todorov:Thesis}. 
To apply it to our HCI purposes, the following three main difficulties need to be dealt with:
First, Todorov's model replicates many phenomena observed in human movement only qualitatively;
there is no known method for adjusting the model to replicate specific experimental data.
Second, the exact movement time needs to be known in advance, which is rarely the case in HCI.
Third, motor control models usually model movement of the human body per se, e.g., movement of the hand as measured through motion capture or a stylus tablet, while the mouse has been avoided.
Mouse pointer movements, however, are modified by sensor characteristics such as mouse sensor rotation and calculations on the microcontroller and in the operating system.
It is unclear whether models that have been developed for understanding natural human (hand) movements are also good models for mouse pointer movements.

In this paper we present an OFC model that addresses all these points.
Based on OFC theory (see Figure~\ref{fig:genmodel}), our two key assumptions are first that control of the system is calculated via \emph{optimization}, i.e., by minimizing a certain cost function. 
Second, the control is obtained in a \emph{feedback} manner, i.e., it depends on the system state.
To provide a simple model to introduce OFC to HCI and the modeling of mouse pointer movements, we make four key simplifications. 

First, following existing literature, we require the cost function that users are assumed to minimize to be \emph{quadratic}.
In pointing tasks, people aim at bringing the end-effector to the target.
For various settings, this has been modeled in OFC literature through {\em quadratic distance costs} that penalize the distance of the end-effector to the target center \cite{Todorov:Thesis,DIEDRICHSEN201031,Qian2013}, see also~\cite{gori18ieee}.  
At the same time, people aim at minimizing their effort and moving smoothly.
The common model for the latter is that users aim to minimize the jerk of the movement~\cite{Flash85}.
Thus, similar to Todorov~\cite{Todorov:Thesis}, we assume the cost function to include terms for penalizing the distance between pointer and target as well as terms to penalize the jerk.

Second, we assume {\em linear dynamics} of the mouse pointer (the \emph{System} block in Figure~\ref{fig:genmodel}).
More precisely, as in Todorov~\cite{Todorov:Thesis}, our system dynamics are described by a second-order lag.

With the third and fourth simplification, we deviate from Todorov~\cite{Todorov:Thesis}: We assume that there are no internal {\em delays} in the model. Moreover, we do not model noise and thus have a \emph{deterministic model}. 
As a result, our approach quantitatively predicts position and velocity of the mouse pointer over time.
In this deterministic setting, fitting the model parameters to the behavior of particular users in a specific task becomes easier.

To summarize, we assume \emph{optimal closed-loop behavior} with respect to a \emph{quadratic cost function} (that penalizes the jerk as well as the distance to the target) and subject to \emph{linear system dynamics} (second-order lag) with {\em no delay} and \emph{no noise}.
These simplifications allow us to solve the optimal control problem using a simple optimal feedback controller, LQR, as explained in the next section.

\section{The Model}

Since mouse sensor data are available in discrete time, we use discrete-time dynamics.
The state of the system is given by a vector~$x_n$ that includes the position and velocity of the virtual mouse pointer.
The user controls the mouse pointer by a force~$u_n$, which influences the state~$x_n$.
Both are given at the discrete time steps $n\in\{1,\dots,N\}$ up to some final $N\in \mathbb{N}$.
The next state $x_{n+1}$ 
depends on the current state $x_n$ and control $u_n$, as described by
\begin{equation}\label{eq:LQR Control1}
	x_{n+1}=Ax_{n}+Bu_{n},
\end{equation}
where the initial state $x_1$ is given.
In this, the matrix $A$ describes how the system, e.g., the mouse pointer dynamics described by a second-order lag, evolves when no control is exerted.
The matrix $B$ describes how the control influences the system. 
In this paper we look at 1D pointing tasks, in which the mouse can only be moved horizontally.
Thus, in our case, the state $x_n$ encodes the horizontal position and velocity of the pointer, denoted by $p_n\in\R$ and $v_n\in\R$, respectively, as well as a target position $T\in\R$ for technical reasons (in order to later be able to compute the distance to the target), i.e.,
\begin{equation}\label{eq:x_n}
	x_n := \left(p_n,v_n,T\right)^\top.
\end{equation}
This model can easily be extended to 2D or 3D pointing tasks by augmenting $x_{n}$ and $u_{n}$ with the respective components for the additional dimensions.

\begin{figure}[!t]
		\subfloat[Mouse pointer model with spring and damper]{\includegraphics[width=\linewidth]{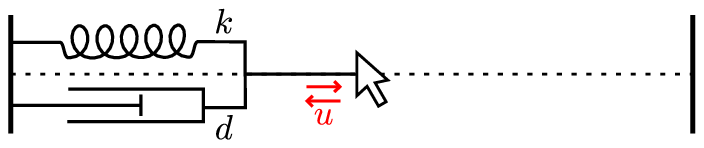} \label{fig:2OL_mouse} }\\
		\subfloat[Control-flow diagram]{	
		\centering
		\resizebox{\linewidth}{!}{
			\begin{tikzpicture}[auto, thick, node distance=1.5cm, >=triangle 45,inner sep=2mm]
			\draw
			node [input] (input1) {} 
			node [input,right of=input1] (plhold){} 
			node [sum, right of=input1,align=left,node distance=2cm] (suma1) {}
			
			node [block, right of=suma1,node distance=2cm] (inte1) {\inte}
			node [split, right of=inte1] (split1) {}
			node [block, right of=split1] (inte2) {\inte}
			node [split, right of=inte2] (split2) {}
			node [output,right of=split2] (output1) {} 
			node [gain, below of=inte1,node distance=0.85cm] (gaind) {$d$}
			node [gain, below of=inte2,node distance=1.35cm] (gaink) {$k$}
			node [input,left of=gaind, node distance=2.65cm] (plhold3) {}
			;
			\draw[->](input1) --  node[near start] {$u(t)$}
			node[pos=0.9]{$+$}(suma1);
			\draw[->](suma1) -- node[near start]{$\ddot{y}(t)$} (inte1);
			\draw[-](inte1) -- node {$\dot{y}(t)$} (split1);
			\draw[->](split1) -- node{} (inte2);
			\draw[->](split1) |- node{} (gaind);
			\draw[->](inte2) -- node {} (output1);
			\draw[-](inte2) -- node {$y(t)$} (split2);
			\draw[->](split2) |- node{} (gaink);
			\draw[->](gaind) -| node[pos=0.9,right]{$-$} (suma1);
			\draw[-](gaink) -|  (plhold3);
			\draw[->](plhold3) -- node[pos=0.9,left]{$-$} (suma1);
			\draw[->](split2) -- node{} (output1);
			\end{tikzpicture}
			
		}
	}
		\caption{Illustrations of the second-order lag~\eqref{eq:2ol-continuous}.~\label{fig:2OL}}

\end{figure}

As a model for the mouse pointer dynamics we use the second-order lag,
as depicted in Figure~\ref{fig:2OL}(a).
The parameters of the model are the stiffness of the spring $k>0$ and the damping factor $d>0$.
The mass is a redundant parameter and does not change the qualitative behavior of the model.
We therefore set it to~1.
In continuous time, we denote the position of the mouse pointer as~$y(t)$, and its first and second derivatives with respect to time (i.e., velocity and acceleration) as~$\dot{y}(t)$ and~$\ddot{y}(t)$, respectively. 
The behavior is then described by the second-order lag equation
\begin{equation}\label{eq:2ol-continuous}
	\ddot{y}(t) = u(t)-ky(t)-d\dot{y}(t),
	\tag{2OL}
\end{equation}
cf.\ Figure~\ref{fig:2OL}(b).
We derive a discrete-time version of~\eqref{eq:2ol-continuous} via the forward Euler method, with a step size of $h=2ms$, where the two milliseconds correspond to the mouse sensor sampling rate.
From this, we obtain the matrices $A$ and $B$ for~\eqref{eq:LQR Control1} as
\begin{equation}\label{eq:dynamics-AB}
	A := 
	\begin{pmatrix}
		1	& h	& 0 \\
		-hk	& 1-hd	& 0 \\
		0	& 0	& 1
	\end{pmatrix},
	\quad
	B := 
	\begin{pmatrix}
		0 \\
		h \\
		0
	\end{pmatrix}.
\end{equation}
This process is similar to the one used by Todorov~\cite{Todorov:Thesis}.

Next, we design the cost function $J_{N}$ that we assume the user to minimize, based on our modeling assumptions. 
We want to penalize the jerk and the distance to the target.
Ideally, no distance costs should occur within the target, which is a box with target width $W$.
Unfortunately, this is infeasible in our LQR setting, where we need cost terms to be quadratic. 
To circumvent this limitation, we construct the distance costs such that we have lower costs inside the target and higher costs outside.
At time step $n$, the {\em remaining} distance to the target is given by $D_n:=|p_n-T|$, and we define the resulting distance costs as the square of that:
\begin{equation}
	D_{n}^{2}=(p_{n}-T)^{2}.
	\label{eq:poserror}
\end{equation}

As in Todorov~\cite{Todorov:Thesis}, the jerk in our case corresponds to the derivative of the control~$u$.
We call~$j_n$ the approximation of the jerk at time step~$n$ obtained by backward differences, i.e., ${j_n := (u_n-u_{n-1})/h \approx \dot{u}_n}$.
We square this term to get positive values only.
A weight factor $r>0$ describes how important the jerk is compared to the positional error~\eqref{eq:poserror}.
Thus, our jerk costs are
\begin{equation}
	rj_n^2 = r\left(\frac{u_{n}-u_{n-1}}{h}\right)^2.
	\label{eq:jerkcost}
\end{equation} 
Formally, this approach requires a value~$u_0$ to be chosen, which we will explain later.

Our overall cost function~$J_N$ will depend on different summations of the distance costs~\eqref{eq:poserror} and the jerk costs~\eqref{eq:jerkcost} over multiple time steps.
In order to design a cost function~$J_N$ that explains user behavior best, we explore three different cost functions of this type later in the paper.

In conclusion, we model the process of pointing through the following optimal control problem:
\begin{equation}\label{eq:ocp}
	\min_{x,u}\ J_N(x,u) \quad\text{subject to}\quad x_{n+1}=Ax_{n}+Bu_{n},
	\tag{OCP}
\end{equation}
for a given initial control $u_0$ and initial state $x_1$, and where the matrices $A$ and $B$ are given by \eqref{eq:dynamics-AB} and the function $J_N$ is some summation of \eqref{eq:poserror} and \eqref{eq:jerkcost} over multiple time steps.

We assume that the user computes the \emph{optimal} control $u_n$, which we denote by $u^*_n$, in a feedback manner.
It has been proven that for these kinds of problems the optimal control $u^*_n$ depends linearly on the state~\cite{dorato-levis71}.
In our case, the optimal control $u_{n}^{*}$ can be calculated simply by multiplying a matrix $-K_{n}$ with the state $x_{n}$, extended\footnote{This extension is required in order to penalize the jerk as in~\eqref{eq:jerkcost}.} by the previous control $u^*_{n-1}$: 
\begin{equation}
	u_{n}^{*}=-K_{n}
	\begin{pmatrix}
		x_{n} \\ u^*_{n-1}
	\end{pmatrix}.
\end{equation}
The matrix~$K_n$ is called the {\em feedback gain} at time step~$n$.
It can be computed directly, given the matrices~$A$, describing the mouse pointer dynamics, and $B$, describing how control influences the mouse pointer, and the cost function~$J_N$. 
This is done by solving the appropriate \emph{Discrete Riccati Equation}, see~\cite[Theorem 7]{Todorov:Thesis}.

The main question now is whether this optimal feedback corresponds to users' behavior, i.e., if our approach is suitable to describe pointing tasks.
For this purpose, we note that there are several free parameters that we can choose: the spring stiffness~$k$, the damping~$d$, and the jerk weight~$r$. 
The goal is to choose these parameters such that users' behavior is approximated best.

\section{Parameter fitting}
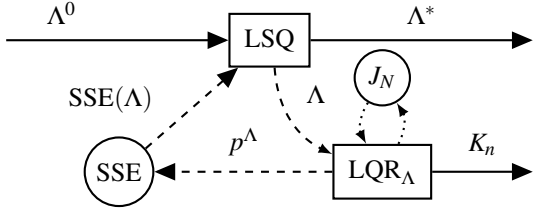
\begin{figure}
	\centering
	\begin{tikzpicture}[auto, thick, node distance=1.5cm, >=triangle 45,inner sep=2mm]
	\draw
	node [input] (input) {} 
	node [block, right of=input1, node distance=3.5cm] (LSQ) {LSQ}
	node [output,right of=LSQ, node distance=3.5cm] (output) {} 
	node [output,below of=output, node distance=1.75cm] (output2) {} 
	node [block, left of=output2, node distance=2cm] (LQR) {$\text{LQR}_\Lambda$}
	node [sum, above of =LQR , inner sep=1mm,node distance=1.25cm] (J) {$J_N$}

	node [sum, left of=LQR, inner sep=1mm, node distance=3.5cm] (SSE) {SSE}
	
	;
	\draw[->](input) --  node[near start] {$\Lambda^0$} (LSQ);
	\draw[->](LSQ) --  node[] {$\Lambda^*$} (output);
	\path  [-latex] (LSQ) edge [bend right,  dashed] node[] {$\Lambda$} (LQR);
	\draw[->, dashed](LQR) --  node[above] {$p^\Lambda$} (SSE);
	\draw[->, dashed](SSE) --  node[near start] {$\text{SSE}(\Lambda)$} (LSQ);
	\draw[->](LQR) --  node[] {$K_n$} (output2);
	\path  [-latex] (J) edge [bend right,  dotted] (LQR);
	\path  [-latex] (LQR) edge [bend right,  dotted] (J);
	
	\end{tikzpicture}
	\caption[Caption for alg]{Starting with an initial parameter set $\Lambda=\Lambda^0$, the least squares (LSQ) algorithm obtains the sum squared error value (SSE) for the currently considered parameter set $\Lambda$.
	To do this, it calls $\text{LQR}_\Lambda$, which sets up the respective optimal control problem \eqref{eq:ocp} and obtains the corresponding optimal feedback gain $K_n$.
	The resulting position time series~$p^\Lambda$ is used to compute $\text{SSE}(\Lambda)$, which is transmitted back to LSQ.
	As an LSQ algorithm, we use MATLAB's nonlinear least squares algorithm \texttt{lsqnonlin}, which uses a gradient-based search method to obtain the next set of parameters $\Lambda$ until it convergences to an optimal parameter set $\Lambda^*$ with minimal SSE.
	Finally, $\Lambda^*$ is returned along with the respective optimal feedback gain matrices $K_n$.}~\label{fig:alg}
\end{figure}

In contrast to the non-deterministic E-LQG model of Todorov~\cite{Todorov:Thesis}, one main strength of our deterministic model is that we can imitate user data without information about the end time of the movement. 
In addition, the calculation of optimal parameters is simplified by eliminating uncertainties. 
In this way, our model can replicate the behavior of a particular user in a particular task. 
To this end, we need to fit the free parameters $k$, $d$, and $r$, to the data.
We denote the set of these parameters by $\Lambda = \{k, d, r\}$. 
The goal is to find the optimal set, $\Lambda^*$, in the sense that our model, with parameters $\Lambda^*$, yields a pointer trajectory that is as similar as possible to that of the user.
To achieve this, we measure the difference between the model trajectory $p^{\Lambda}$ and the user trajectory $p^{\text{USER}}$ using the sum squared error (SSE):
\begin{equation}\label{eq:SSE-obj-fct}
	\text{SSE}(\Lambda)=\sum_{n=1}^{N} \left(p_{n}^{\Lambda} - p_{n}^{\text{USER}}\right)^{2}.
\end{equation}
We then apply the least squares (LSQ) algorithm depicted in Figure~\ref{fig:alg} to find the optimal parameter set~$\Lambda^*$ minimizing~\eqref{eq:SSE-obj-fct}.

Least-squares-based algorithms may converge to local minima and not find a global minimum.
Therefore, we execute the whole fitting process several times for randomly chosen starting parameter sets~$\Lambda^0$.
According to our simulations, 100 of such sets sufficed to provide results that would not improve further by iterating on more starting parameter sets.

\section{Pointing task and dataset}
To evaluate our model, we use the {\em Pointing Dynamics Dataset}.
Task, apparatus, and experiment are described in detail in \cite{mueller17}.
The dataset contains the mouse trajectory for a reciprocal pointing task in 1D for ID 2, 4, 6, and 8.

Pointing movements almost always start with a reaction time, in which velocity and acceleration of the pointer are close to zero.
In real computer usage, the user usually takes some time to decide whether to move the mouse and to locate the target before initiating the movement.
Therefore, one could speak of the movement beginning once the acceleration of the pointer reaches a certain threshold.

In the Pointing Dynamics Dataset we use, the trial started immediately when the previous trial was finished, i.e., after the mouse click, not when the user initiated the next movement.
This results in a considerable variation in reaction times.
Since some variants of our approach as well as the methods from the literature we use for comparison cannot properly handle reaction times, in each trial we ignore the data before the user starts moving. 
To be exact, we drop all frames before the acceleration reaches $0.5\%$ of its maximum/minimum value (depending on the movement direction) for the first time in each trial.

Moreover, we ignore user mistakes by dropping the failed and the following trial.
From all other trials of all participants and all tasks -- 7732 trajectories in total -- we have removed another 30 for which the optimally fitted damping parameter~$d$ was an outlier (more than three standard deviations from the mean).
This was necessary due to numerical instabilities that occurred for these parameters, leading to erroneous calculations of the optimal control.
All remaining 7702 trajectories are used in the later evaluation.

We use the raw, unfiltered position data in our parameter fitting process to avoid artifacts.
The dataset also contains derivatives of user trajectories, which were computed by differentiating the polynomials of a Savitzky-Golay filter of degree 4 and frame size 101 \cite{mueller17}.
We use this (filtered) data only for the computation of the reference control $u_{0}$ (see the next chapter) and for illustration purposes.

For the following plots, unless stated otherwise, we display one certain representative user trajectory, namely the $21^{\text{st}}$ movement to the right of participant 1 for the ID 8 task with 765px distance and 3px target width.
For comparison and validation, the plots of all 7702 trajectories are provided in the supplementary material.

\section{Iterative design of the cost function}
In this section we describe the iterative design of our cost function~$J_N$ that is utilized in the algorithm depicted in Figure~\ref{fig:alg}.
The three resulting approaches are denoted by \mbox{2OL-LQR} with the corresponding numbering.

\subsection{First Iteration: Distance Costs at Endpoint (2OL-LQR$_1$)}

In our first iteration we use a cost function similar to the one used by Todorov~\cite{Todorov:Thesis} for the E-LQG model.
In this function, jerk costs occur at every step.
Distance costs, however, only occur in the time step in which the mouse is clicked (time step~$N$).
In particular, no distance costs occur at other time steps. 
Thus, the cost function is given by 
\begin{equation}\label{eq:Cost Functional2}
	J_{N}(x,u)=D_{N}^2 + r \sum_{n=1}^{N-1}j_n^2,
\end{equation}
where $D_N=|p_N-T|$ is the {\em remaining} distance to the target center at the end of the movement, $r$ is the weight of the jerk, and $j_n = (u_n-u_{n-1})/h$ is the jerk at time step $n$.

The initial pointer position and velocity are set from the data, i.e., $x_{1}=(p_{1}^{\text{USER}},v_{1}^{\text{USER}},T)^\top$.
Although the choice of $u_0$ does not have a direct impact on the system dynamics, the trajectory heavily depends on its value. 
This is due to $j_1$ penalizing the deviation of $u_1$ from $u_0$, which carries over to $j_2$, and so on.\footnote{For example, setting $u_0=0$ might result in an implausibly high acceleration at the start of the movement, similar to 2OL-Eq.}
We define~$u_{0}$ such that if the first control $u_1$ coincides with~$u_0$, the model will replicate the initial acceleration from the data~$a_{1}^{\text{USER}}$, i.e., \mbox{$u_0=kp_{1}^{\text{USER}}+dv_{1}^{\text{USER}}+a_{1}^{\text{USER}}$}.

\begin{figure}
	\centering
	\includegraphics[width=0.9\columnwidth]{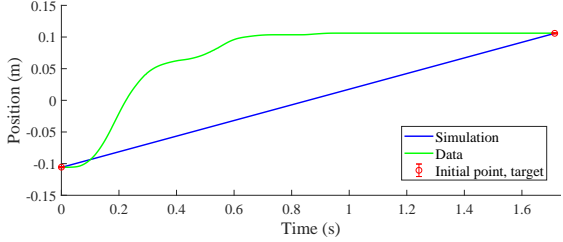}
	\caption{First iteration (2OL-LQR$_{1}$): Using a cost function similar to the one proposed by Todorov
          results in the model (blue) not replicating the data (green) well.
          }
        \label{fig:var1}
\end{figure}

The approach of using cost function~\eqref{eq:Cost Functional2} suffers from two major problems.
First, as illustrated in Figure~\ref{fig:var1}, the generated trajectories do not fit our data.
In particular, the target is reached only at exactly the time of the mouse click.
In contrast, our data shows that for high IDs, the users reach the vicinity of the target much earlier and then spend considerable time with small corrective submovements close to the target.
The reason for this different behavior is that the cost function~\eqref{eq:Cost Functional2} sets the incentive to settle at the target only at the final time step~$N$, while the jerk is penalized in every time step. 

The second problem is that the cost function must include the exact time of the mouse click a priori.
This makes the cost function very difficult to use for the simulation of human behavior in pointing tasks, if we cannot or do not want to prescribe a specific clicking time.

Hence, we propose a slightly modified cost structure in the LQR algorithm to take these considerations into account.

\subsection{Second Iteration: Summed Distance Costs (2OL-LQR$_2$)}
Both issues of the first iteration can be attributed to the fact that the remaining distance to target is only penalized at the time of the mouse click.
Hence, we now penalize both the jerk and the distance between pointer position and target during the whole movement.
Having summed costs over the entire movement is a standard approach in optimal control for such tracking tasks \cite{chan-maille75}.
Our new cost function is
\begin{equation}\label{eq:Cost Functional3}
	J_{N}(x,u)=D_{N}^2+\sum_{n=1}^{N-1}\left(D_{n}^2+rj_n^2\right),
\end{equation}
where $D_n=|p_n-T|$ is the {\em remaining} distance to the target center after time step $n$.
This changes the meaning of $N$: Instead of being the exact clicking time, it can now be interpreted as the maximum time allowed for the task.
Thus, it is now much less important to set $N$ accurately.

Optimal solutions of this approach with respect to the new cost function~\eqref{eq:Cost Functional3} approximate most of the considered user trajectories well, and much better than 2OL-LQR$_1$, cf.\ Figure~\ref{fig:var2}.

\subsection{Third Iteration: Reaction Time (2OL-LQR$_3$)}
As explained in the dataset section, we prefer to model only the movement itself, excluding the reaction time. 
Thus, our second iteration does not model reaction time.
In some cases, however, it is desirable to
model it explicitly.
In this section we present an objective function that achieves this. 

To this end, we add a parameter $\delta>0$ that should describe the reaction time. 
Due to our discrete time setting, we introduce $n_{\delta}\in\{1,\ldots,N\}$ as the discrete time step closest to $\delta$.
The idea is to adjust the cost function such that it incentivizes standing still until $n_\delta$, to take reaction time into account.

We achieve this by splitting the cost function in two parts, before and after $n_\delta$.
In the first part, we assume that users are not aware of the target position or have at least not processed all required information for initiating the motion.
In both cases, users should have no interest in changing their control.
Therefore, we do not penalize the distance to the desired position in that time frame and employ a much higher jerk penalization compared to the main movement phase.
More precisely, $r$ is replaced by $f(n)\cdot r$, where $f(n)$ is, for the most part, an approximation of a very large constant $c$, e.g., $c=100000$.\footnote{To aid the LSQ optimization process, we use a smoothed version of the piecewise constant sequence of jerk weights $c\cdot r$ and $r$, i.e.,  $f(n):=(c-1)\exp(\frac{1}{n_{\delta}-1}-\frac{1}{n_{\delta}-n})+1$ for $n \in \{1,\dots,n_{\delta}-1\}$.}
In the second part, i.e., starting from time step $n_\delta$, we use the cost function~\eqref{eq:Cost Functional3} from 2OL-LQR$_2$. 

In total, the cost function of 2OL-LQR$_3$ is
\begin{equation}\label{eq:Cost Functional4}
	J_{N}(x,u)= D_{N}^2 + \sum_{n=1}^{n_{\delta}-1}f(n)r j_n^2 + \sum_{n=n_{\delta}}^{N-1}\left(D_{n}^{2}+rj_n^2\right).
\end{equation}
There are several ways to obtain the reaction time~$\delta$ and thus~$n_\delta$.
One way is to determine it directly from the data, e.g., as the time when the acceleration passes a certain threshold.
Another approach is to include it as an additional parameter to be optimized by the LSQ algorithm. 
We have chosen the latter approach and it works well according to our results.

\section{Results}
In this section we evaluate our main model, 2OL-LQR$_2$, by comparing it to the minimum-jerk model from~\cite{Flash85} (MinJerk) and the second-order lag with equilibrium control from~\cite{mueller17} (2OL-Eq).
We also investigate how the parameters of our model change for different tasks (IDs) and different users.
Finally, we demonstrate the ability of 2OL-LQR$_3$ to model movements including a reaction time.

\subsection{Minimum-Jerk Model by Flash and Hogan (MinJerk)}
Flash and Hogan~\cite{Flash85} show that the minimum-jerk trajectory between two points is a fifth-degree polynomial.
They assume that velocity and acceleration are zero at the start and at the end of the movement, and explain how the parameters of this polynomial can be computed under these conditions.
However, in our dataset, velocity and acceleration are not necessarily zero, neither at the beginning nor at the end of the movement.
Therefore, before we delve into the results, we present the following technique to derive the parameters of the minimum-jerk polynomial under these different conditions. 

\subsubsection{Deriving the MinJerk Polynomial}
In~\cite{Flash85}, the minimum-jerk polynomial is given by
\begin{equation}\label{eq:minjerk}
	p^{\text{MinJerk}}(t)=\sum_{i=0}^{5} c_{i}~\left(\frac{t}{t_{f}}\right)^{i},
\end{equation}
with coefficients $c_0,\ldots,c_5$ and where $t_f$ is the final time of the movement. 
In our discrete-time setting, we evaluate the polynomial only at times $t_{n}=(n-1)h$, $n\geq 1$.
In this case, the final time is given by $t_{f}=(\tilde{N}-1)h$, where $\tilde{N}$ is the last time step\footnote{We specifically do not use $N$ for reasons elaborated below.} and $h$ is the same step size as before.
Thus, the position at time step $n$ is given by
\begin{equation}\label{eq:minjerk-discrete}
	p_{n}^{\text{MinJerk}}=\sum_{i=0}^{5} c_{i}\left(\frac{n-1}{\tilde{N}-1}\right)^{i}.
\end{equation}
The coefficients $c_0,\ldots,c_5$ are computed from the data: $c_0$ is the initial position, i.e., $c_{0}=p_{1}^{\text{USER}}$.
The coefficients $c_1$ and $c_2$ are computed from initial velocity $v_{1}^{\text{USER}}$ and acceleration $a_{1}^{\text{USER}}$.
Since we have to take into account factors arising from differentiation, we arrive at $c_{1}=v_{1}^{\text{USER}}t_f$ and $c_{2}=a_{1}^{\text{USER}}t_f^2/2$. 
The remaining coefficients $c_3, c_4, c_5$ can be computed by solving the system of linear equations
\begin{equation}\label{eq:minjerk-system}
	\begin{pmatrix}
	1 & 1 & 1 \\
	3 & 4 & 5 \\
	6 & 12 & 20
	\end{pmatrix}
	\begin{pmatrix}
	c_{3} \\ c_{4} \\ c_{5}
	\end{pmatrix} =
	\begin{pmatrix}
	p_{t_f}^{\text{USER}} - c_{0} - c_{1} - c_{2}\\ 
	v_{t_f}^{\text{USER}}t_f - c_{1} - 2c_{2} \\ 
	a_{t_f}^{\text{USER}}t_f^2 - 2c_{2}
	\end{pmatrix}, 
\end{equation}
where $p_{t_f}^{\text{USER}}$, $v_{t_f}^{\text{USER}}$, and $a_{t_f}^{\text{USER}}$ are, respectively, the pointer position, velocity, and acceleration at the final time.

\subsubsection{Results for MinJerk}
\begin{figure}
	\centering
	\subfloat[Position Time Series]{\includegraphics[width=1\linewidth]{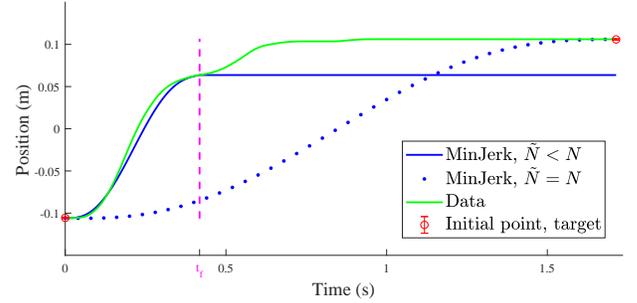} \label{fig:minjerkpos} }
	\\
	\subfloat[Velocity Time Series]{\includegraphics[width=0.5\linewidth]{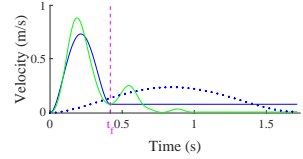}}
	\subfloat[Acceleration Time Series]{\includegraphics[width=0.5\linewidth]{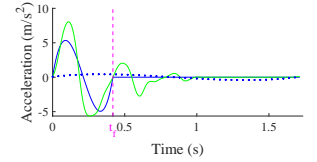}}
	\caption{For the MinJerk model, we have to decide whether we want to model the surge well, but not reach the target (blue solid line with constant continuation after $t_{f}$), or reach the target, but not model the entire movement well (blue dotted line). In this paper we have chosen the former option. In this case $t_f$ is the final time of the surge.}
	\label{fig:minjerk-plots}
\end{figure}

The MinJerk model has been derived from data of an experiment that did not involve any corrective submovements~\cite{Flash85}.
This leaves two possibilities to fit the model to our data, which does show extensive corrective submovements.
If MinJerk is used for modeling the entire movement, i.e., until time step~$N$, the fit is very poor (see Figure~\ref{fig:minjerk-plots}; dotted line).
Instead of a quick movement towards the target with extensive corrective submovements, as in our data, the model predicts a slow, smooth movement, reaching the target only at the time of the mouse click.

Therefore, we use MinJerk for only the first, rapid movement towards the target (the ``surge''). 
Similar to \cite{mueller17}, we determine the end of the surge ($t_f$ in Figure~\ref{fig:minjerk-plots}) from the data as the first zero-crossing in the acceleration time series after the deceleration (for movements to the left: acceleration) phase.
After that, we assume that the pointer does not move.
As illustrated in Figure~\ref{fig:minjerk-plots} (blue solid line), this results in a good fit of the surge phase, at least for movements that exhibit a clear surge phase.
However, the target is not reached, causing a poor overall fit.

In conclusion, MinJerk is a good model for the surge phase but not suitable for describing motions that contain extensive corrective submovements.

\begin{figure}[t]
	\centering
	\subfloat[Position Time Series]{\includegraphics[width=0.5\linewidth]{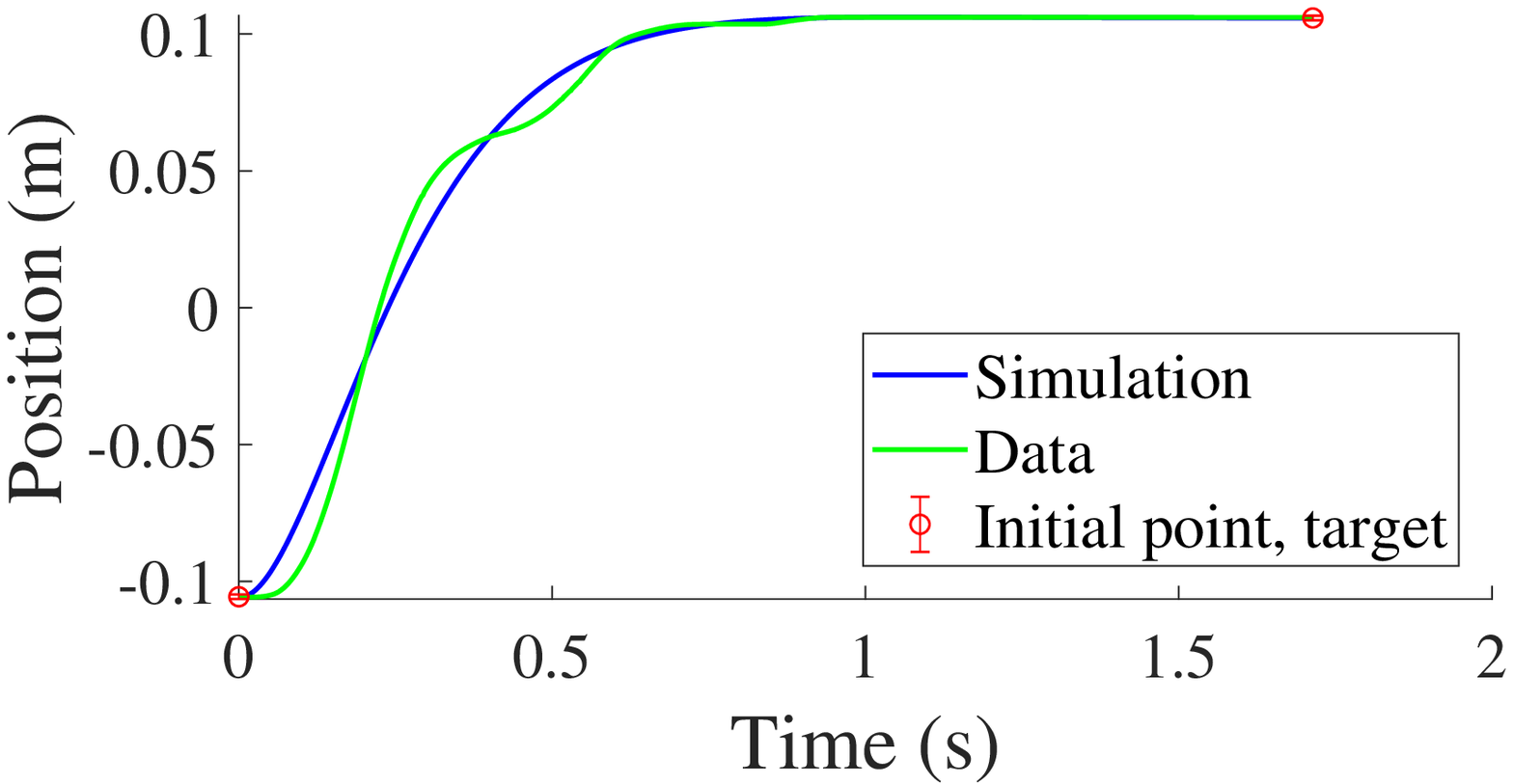}}
	\subfloat[Velocity Time Series]{\includegraphics[width=0.5\linewidth]{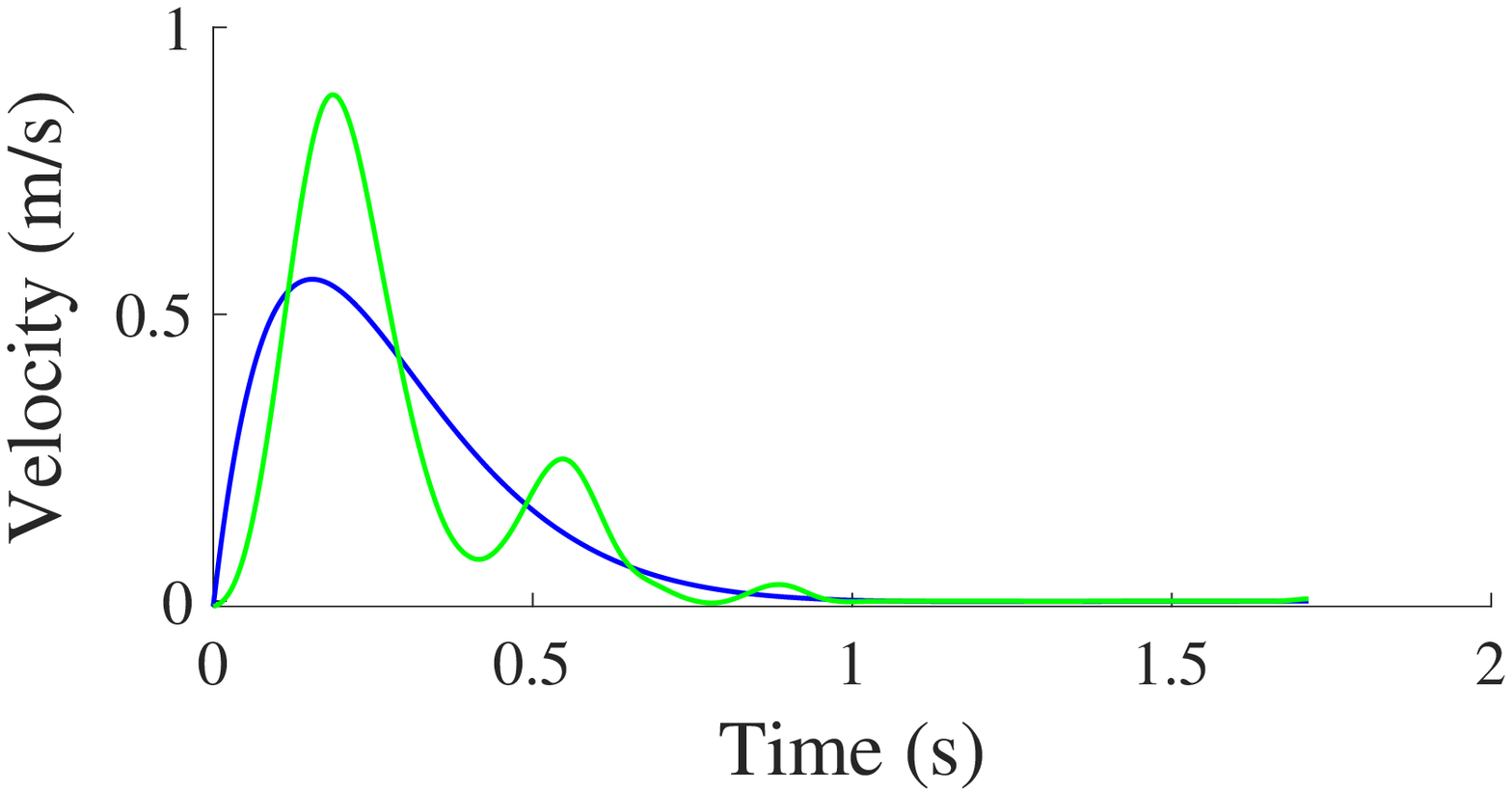}}
	\\
	\subfloat[Acceleration Time Series]{\includegraphics[width=0.5\linewidth]{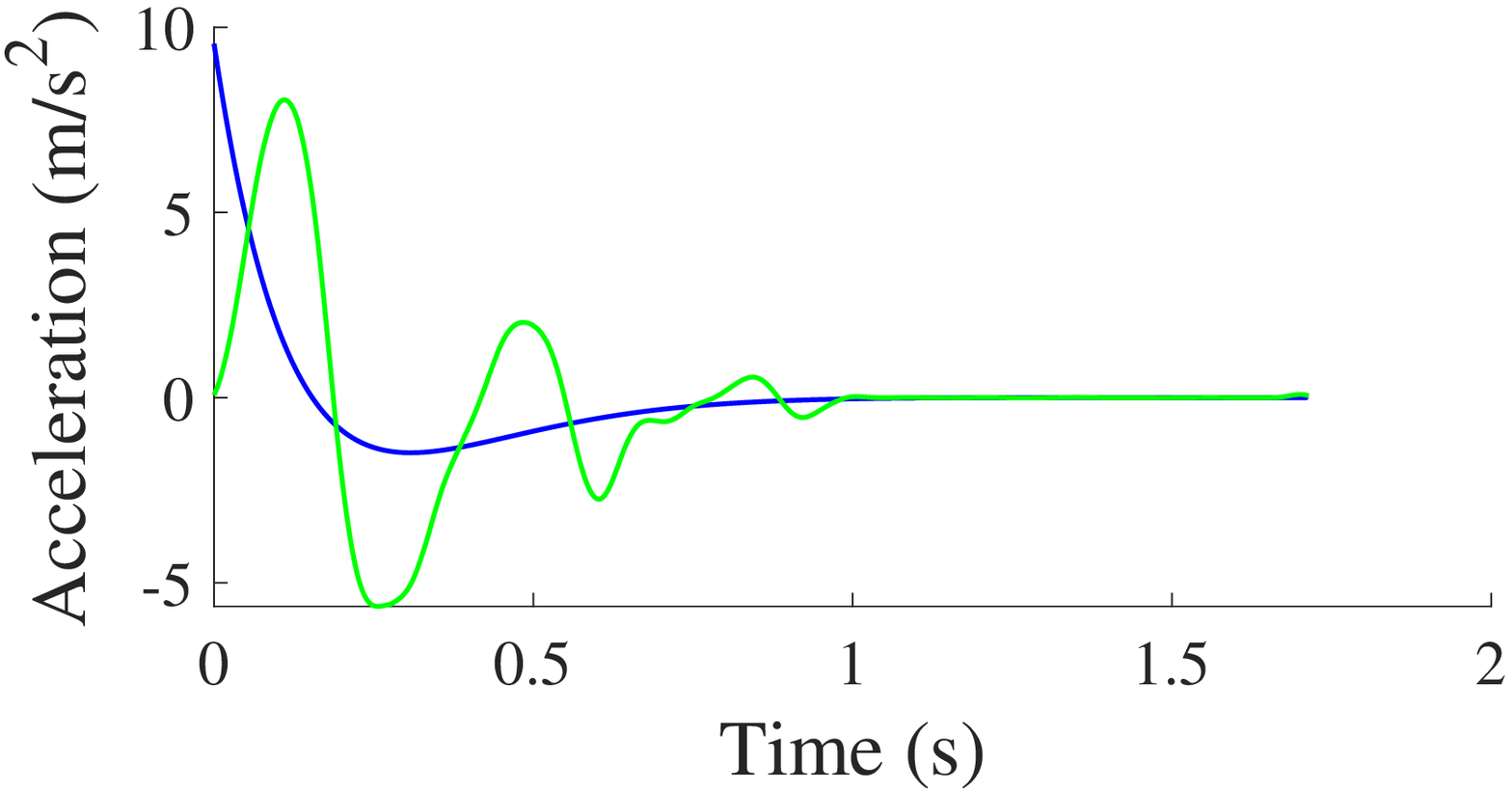}}
	\subfloat[Control Time Series]{\includegraphics[width=0.5\linewidth]{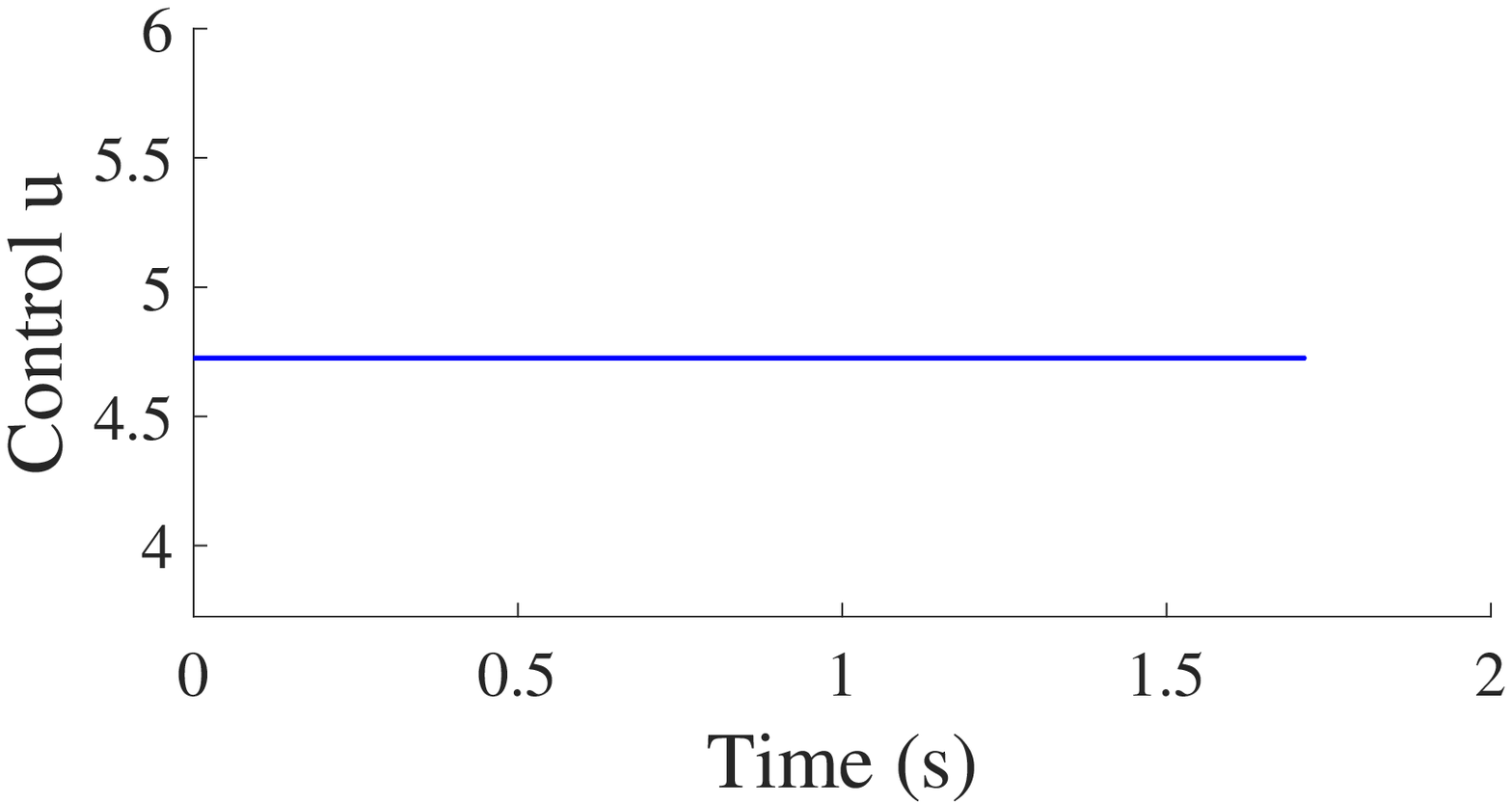}}
	\caption{Due to the constant control, 2OL-Eq yields a much less symmetric velocity and acceleration profile during the surge than the user data.}
	\label{fig:2OL-plots}
\end{figure}

\subsection{Second-order Lag Equilibrium Control (2OL-Eq)}
The 2OL-Eq model is a discrete version of~\eqref{eq:2ol-continuous} with $u \equiv k T$.
It is given by the system dynamics $x_{n+1}=Ax_{n}+Bu_{n}$\linebreak with matrices $A$ and $B$ from~\eqref{eq:dynamics-AB} and initial condition\linebreak $x_{1}=(p_{1}^{\text{USER}},v_{1}^{\text{USER}},T)^\top$.
With this particular choice of control, the pointer moves towards the target $T$ and stays there. 
The target position $T$, together with zero velocity and acceleration, constitutes an equilibrium in this case; hence the name ``equilibrium control''. 
This constant control is the main difference to our approach, in which the control values $u_{n}$ are optimized with respect to some cost function $J_{N}$.

For the 2OL-Eq model, we optimize the spring stiffness $k$ and the damping $d$ with the same parameter fitting process and the same SSE objective function \eqref{eq:SSE-obj-fct} that we use for our 2OL-LQR approach.

The behavior of the 2OL-Eq is shown in Figure~\ref{fig:2OL-plots}. 
Visually, the model captures user behavior well in terms of pointer position, cf.\ Figure~\ref{fig:2OL-plots}(a).
The velocity time series depicted in Figure~\ref{fig:2OL-plots}(b), however, is asymmetric in the 2OL-Eq case, while the user shows a more symmetric, bell-shaped velocity profile.
The biggest difference appears in the acceleration time series.
The user performs a symmetric and smooth N-shaped acceleration. 
In contrast, the acceleration of the 2OL-Eq jumps instantaneously at the start of the movement, and then rapidly declines.
This can be explained with the physical interpretation of the 2OL-Eq as a spring-mass-damper system:
Since $u$ is constant in this model, as the system is released, the spring instantaneously accelerates the system with a force that is proportional to the extension of the spring.
Because human muscles cannot build up force instantaneously~\cite{SchLee05}, this behavior is not physically plausible.

\begin{figure}[!t]
	\centering
	\subfloat[Position Time Series]{\includegraphics[width=0.5\linewidth]{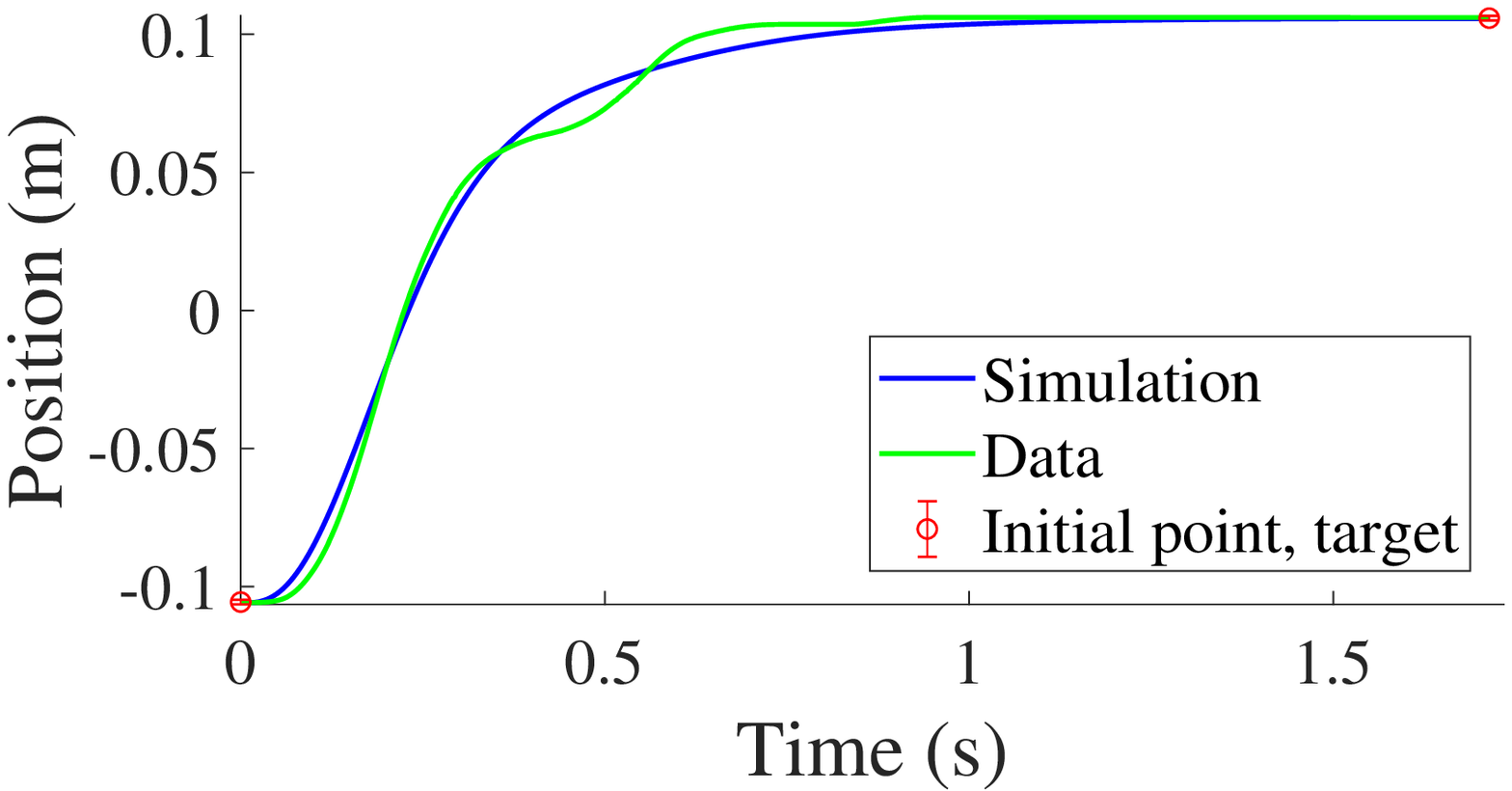}}
	\subfloat[Velocity Time Series]{\includegraphics[width=0.5\linewidth]{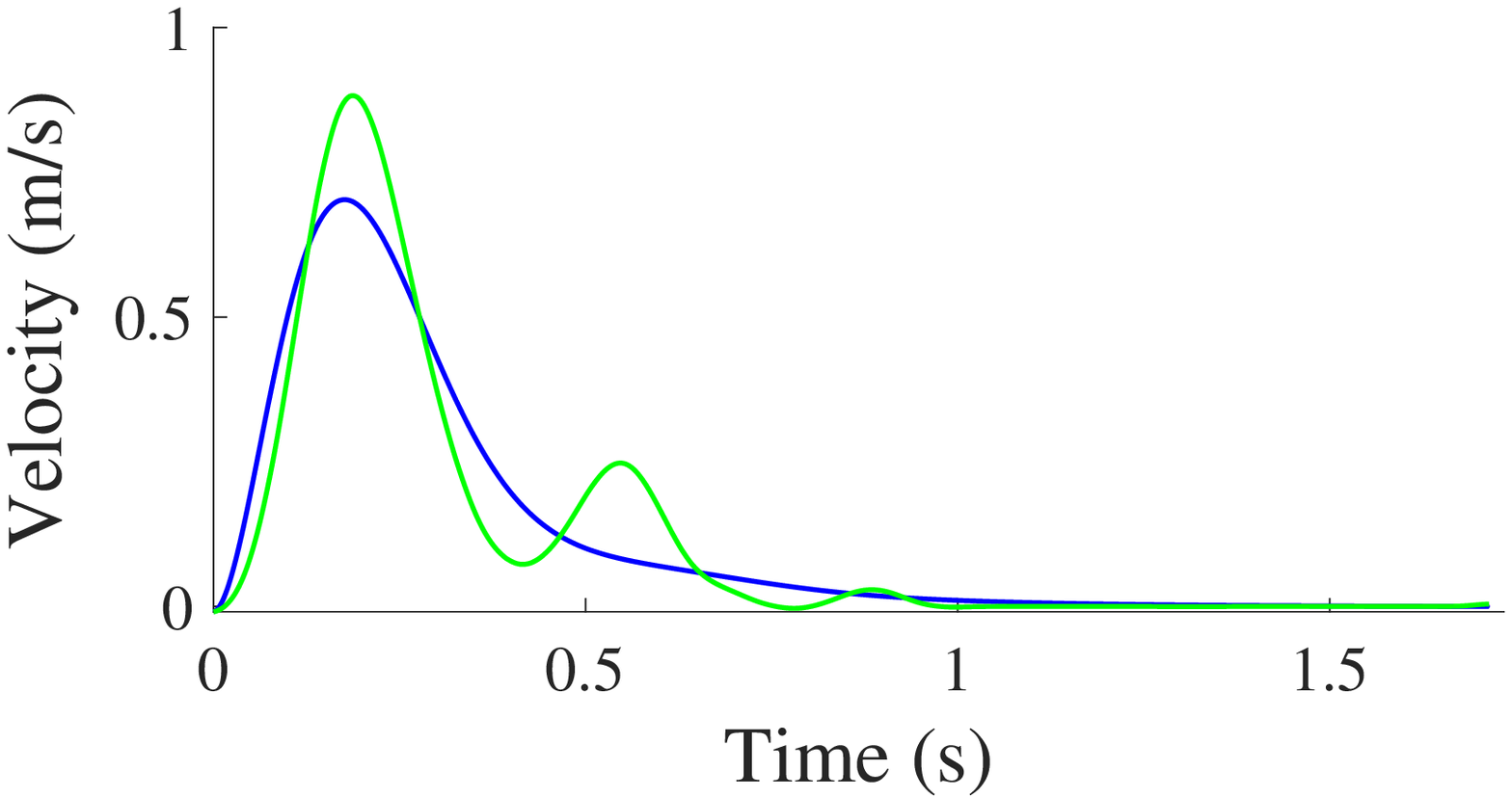}}
	\\
	\subfloat[Acceleration Time Series]{\includegraphics[width=0.5\linewidth]{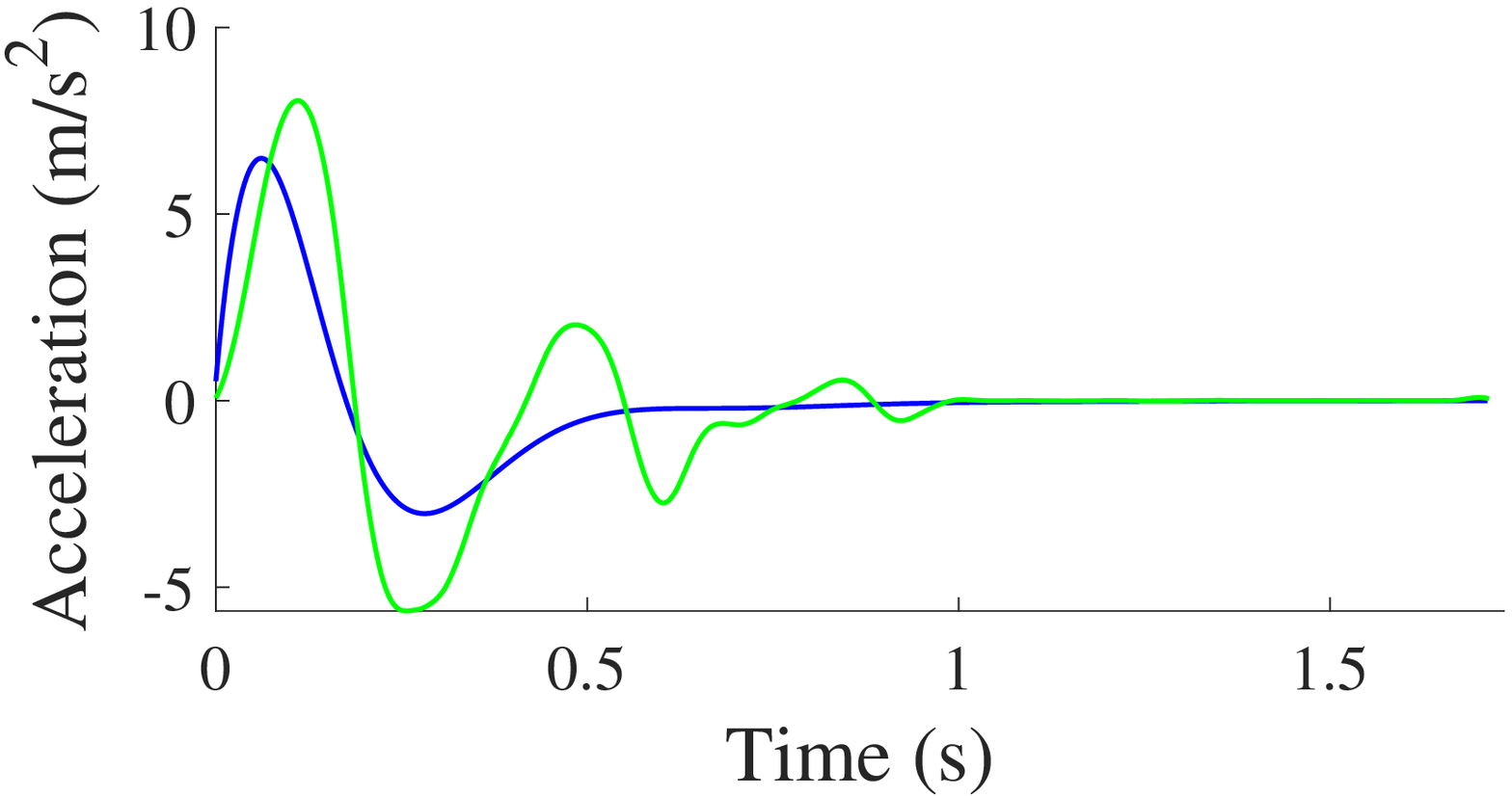}}
	\subfloat[Control Time Series]{\includegraphics[width=0.5\linewidth]{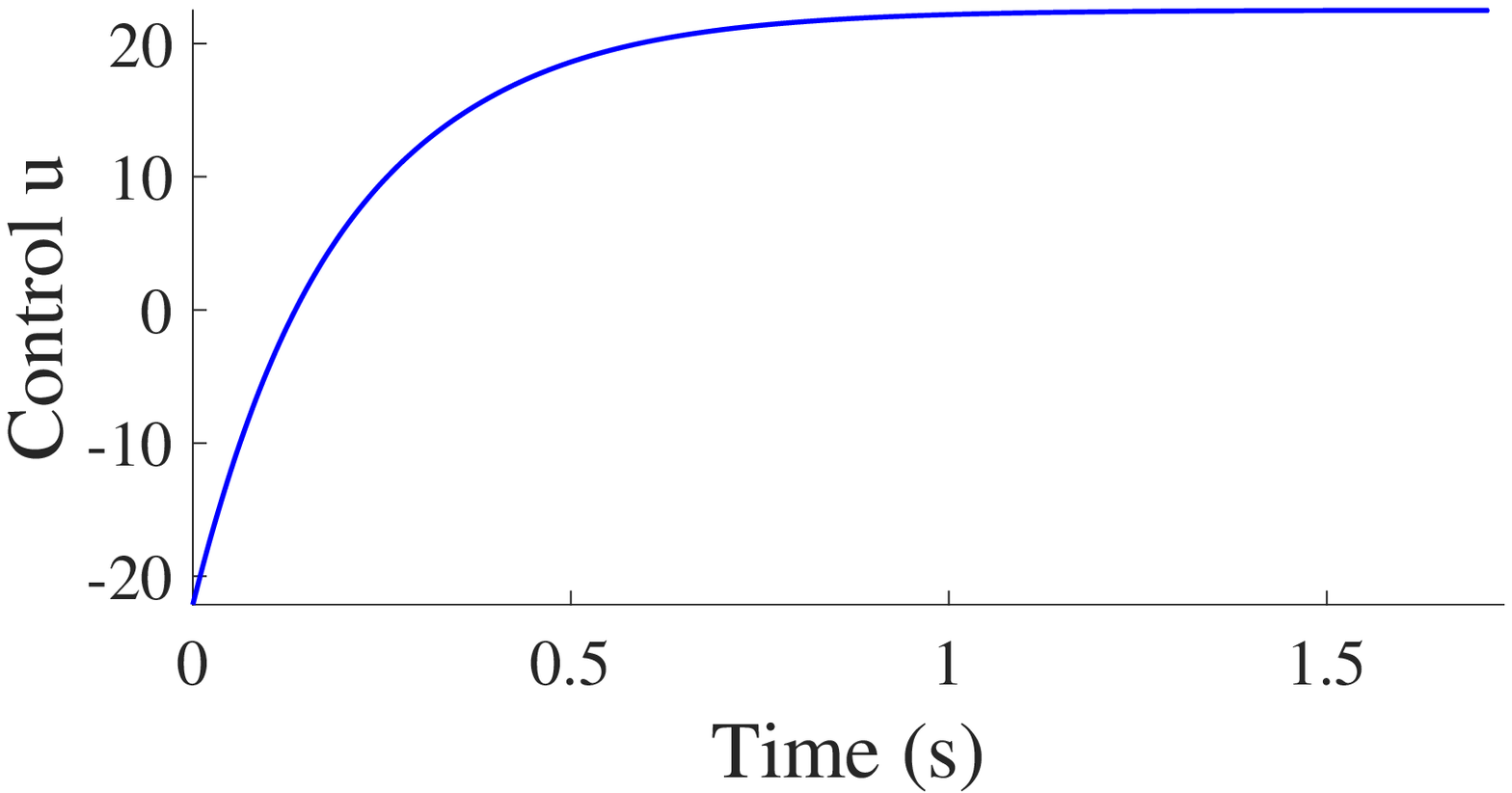}}
	\caption{Our second iteration model 2OL-LQR$_{2}$ models the entire movement well. However, the acceleration in the surge phase is slightly less symmetric than the one of the user.}~\label{fig:var2}
\end{figure}

\subsection{Our Model 2OL-LQR$_2$ vs. MinJerk and 2OL-Eq}
\subsubsection{Qualitative Comparison}
For the qualitative comparison, we performed a visual analysis of model behavior on the entire dataset. 
Although in the figures we illustrate a particular movement of a specific participant, we recall that the behavior is representative and the plots of all 12 participants and all 4 IDs are provided as supplementary material. 

The behavior of our model 2OL-LQR$_2$ is shown in Figure~\ref{fig:var2}.
Overall, the model approximates the position rather well over the entire movement, cf.\ Figure~\ref{fig:var2}(a).
Corrective submovements, which start at around $t=0.4s$, are not replicated well by any of the three models (see Figures~\ref{fig:minjerk-plots}, \ref{fig:2OL-plots}, and~\ref{fig:var2}). 
Our model slightly underestimates the maximum velocity and the velocity profile is less symmetric than the data. 
Similar effects can be observed in the acceleration, see Figure~\ref{fig:var2}(c).

Compared to MinJerk, our model 2OL-LQR$_2$ explains the surge phase similarly well, while not quite capturing the symmetry observed in many acceleration time series as the one depicted in Figures~\ref{fig:minjerk-plots}, \ref{fig:2OL-plots}, and \ref{fig:var2}.\footnote{There are some cases in which asymmetric acceleration time series do occur. Our model 2OL-LQR$_2$ is able to approximate these profiles reasonably well and is not limited to, e.g., an N-shaped acceleration profile, as is the case with MinJerk.}
However, as a major improvement compared to MinJerk, 2OL-LQR$_{2}$ captures the entire movement, not just the surge phase.
We emphasize that MinJerk is given the end point of the surge, as well as position, velocity and acceleration at that point, while our model is not given that information.

Compared to 2OL-Eq, our model captures position, velocity, and acceleration much better.
The reason for this is that, in contrast to 2OL-Eq, the control time series shown in Figure~\ref{fig:var2}(d) is not constant but changes over time.
This often leads to a more N-shaped acceleration time series and a more bell-shaped velocity time series, as predicted by Flash and Hogan~\cite{Flash85} and in many cases confirmed by our data. 

\begin{figure}[!t]
	\centering
	\subfloat[Position Time Series]{\includegraphics[width=0.8\linewidth]{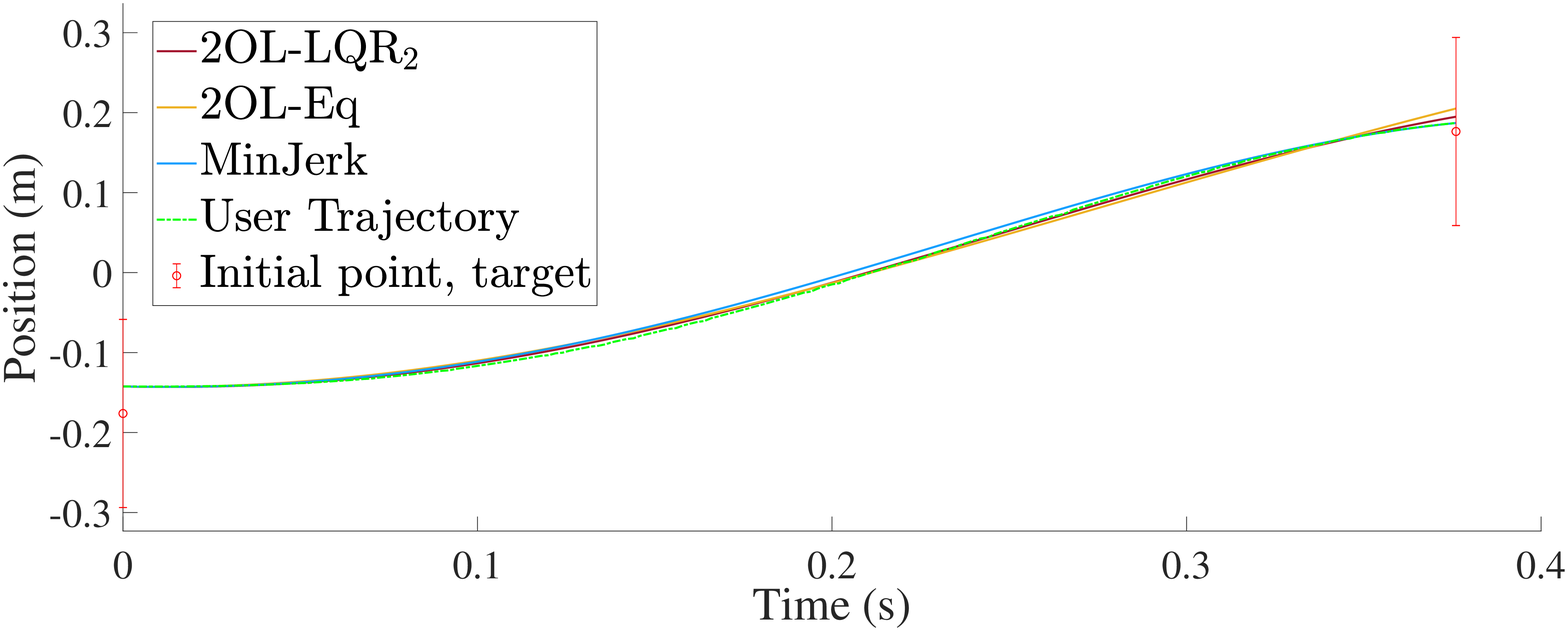} \label{fig:ID2-comparison-position} }
	\\
	\subfloat[Velocity Time Series]{\includegraphics[width=0.5\linewidth]{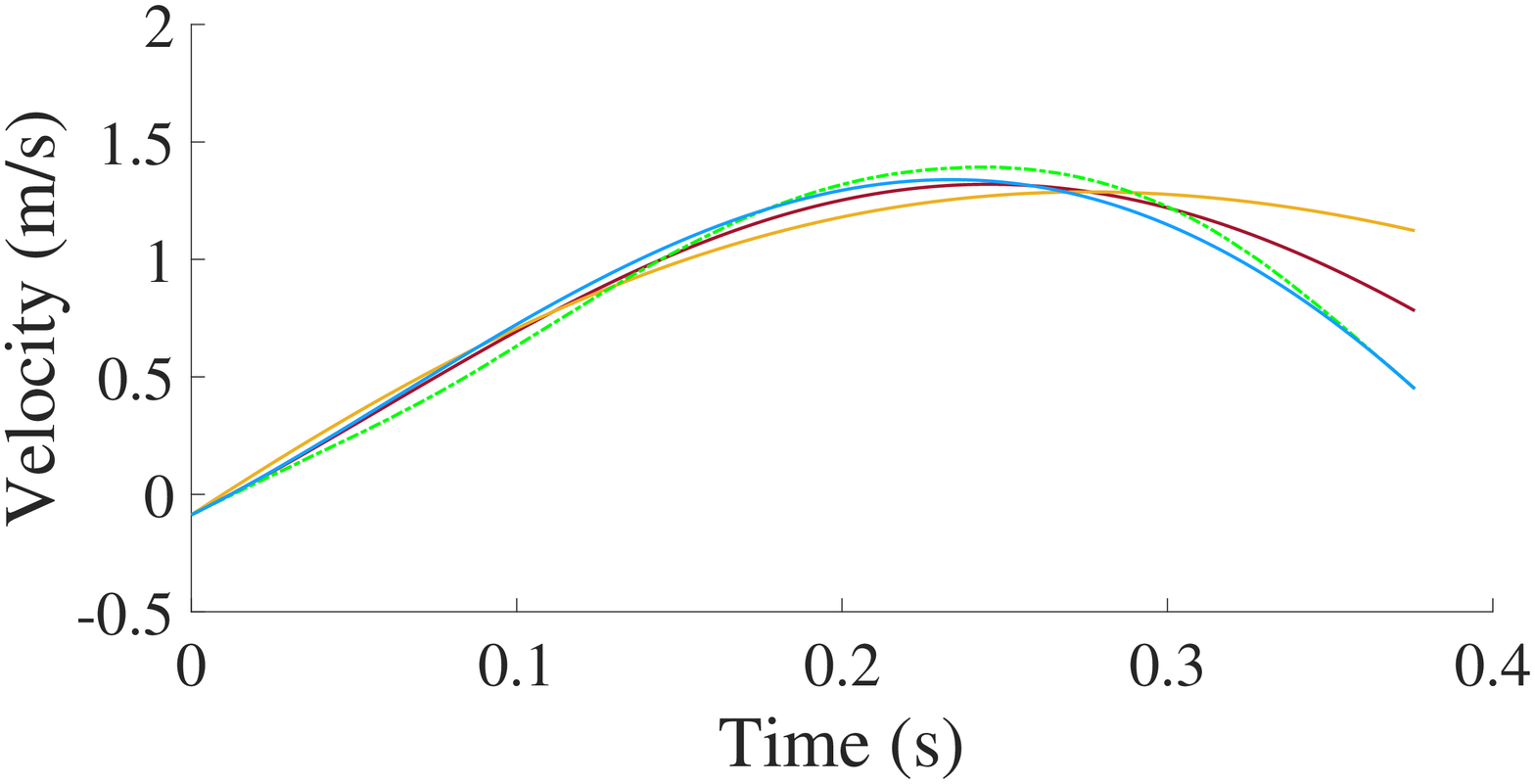}}
	\subfloat[Acceleration Time Series]{\includegraphics[width=0.5\linewidth]{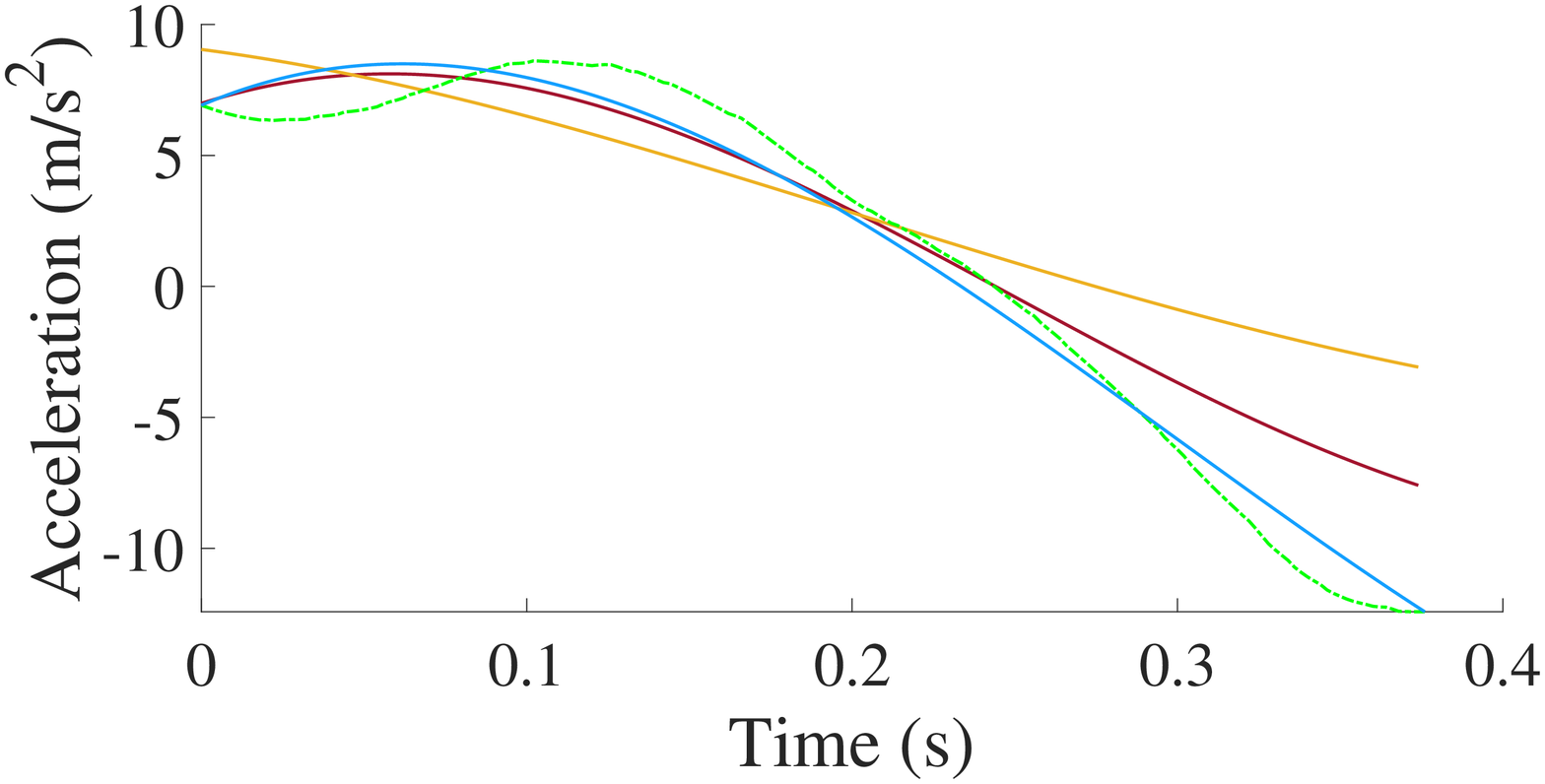}}
	\caption{ID~2 tasks without a correction phase are well approximated by each of the three considered models (here: Participant 1, 1275px distance, 425px target width, 35$^{\text{th}}$ movement to the right).} 
	\label{fig:ID2-comparison-plots}
\end{figure}

ID~2 tasks play a special role, as they (usually) do not involve corrective submovements, see Figure~\ref{fig:ID2-comparison-plots}.
In this case, all three models match the position data.
Visible differences in the fit appear in the velocity and acceleration data.

\subsubsection{Quantitative Comparison}
\begin{figure}
	\centering
	\subfloat[SSE]{\includegraphics[width=0.5\linewidth]{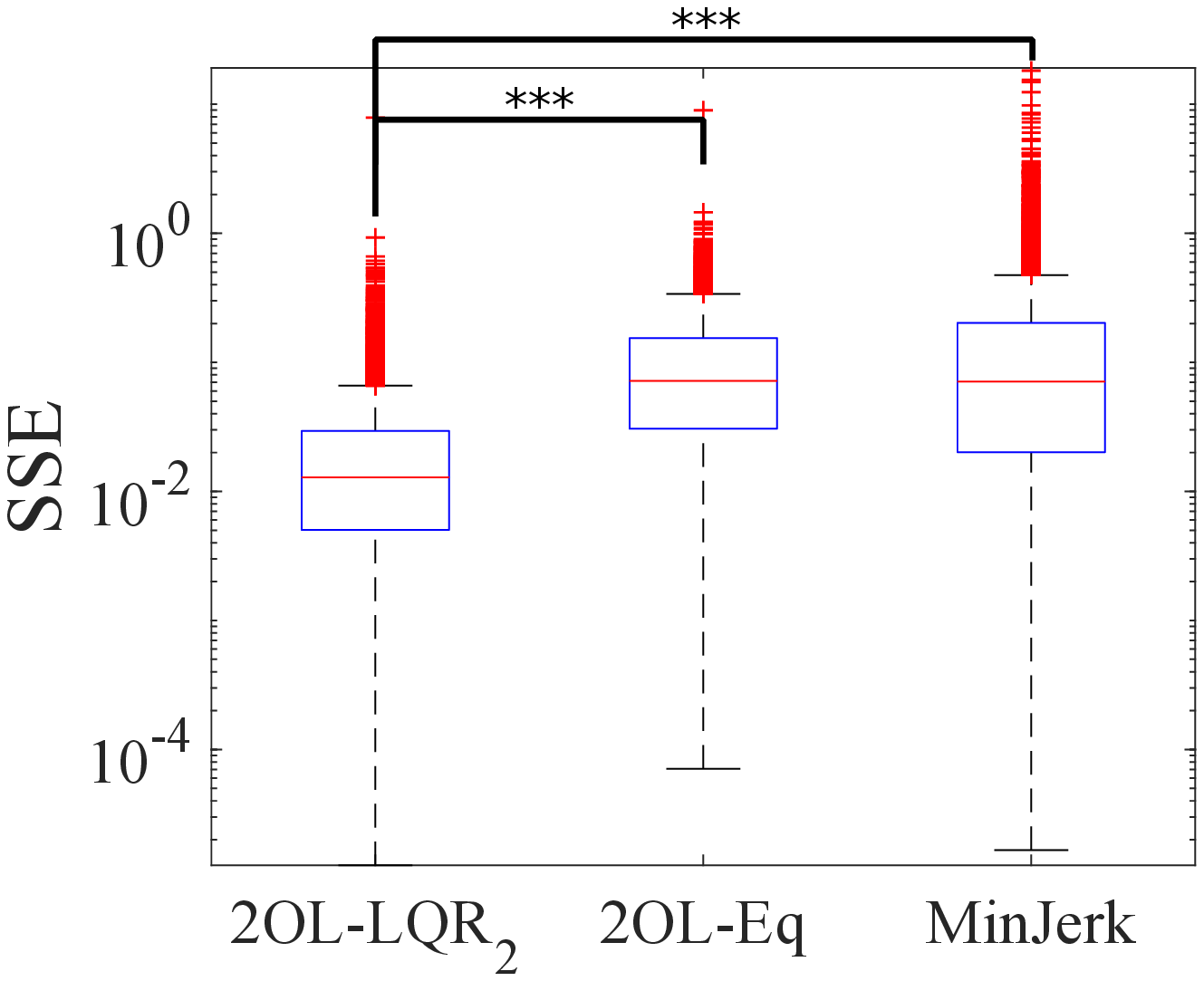}}
	\subfloat[Maximum Error]{\includegraphics[width=0.5\linewidth]{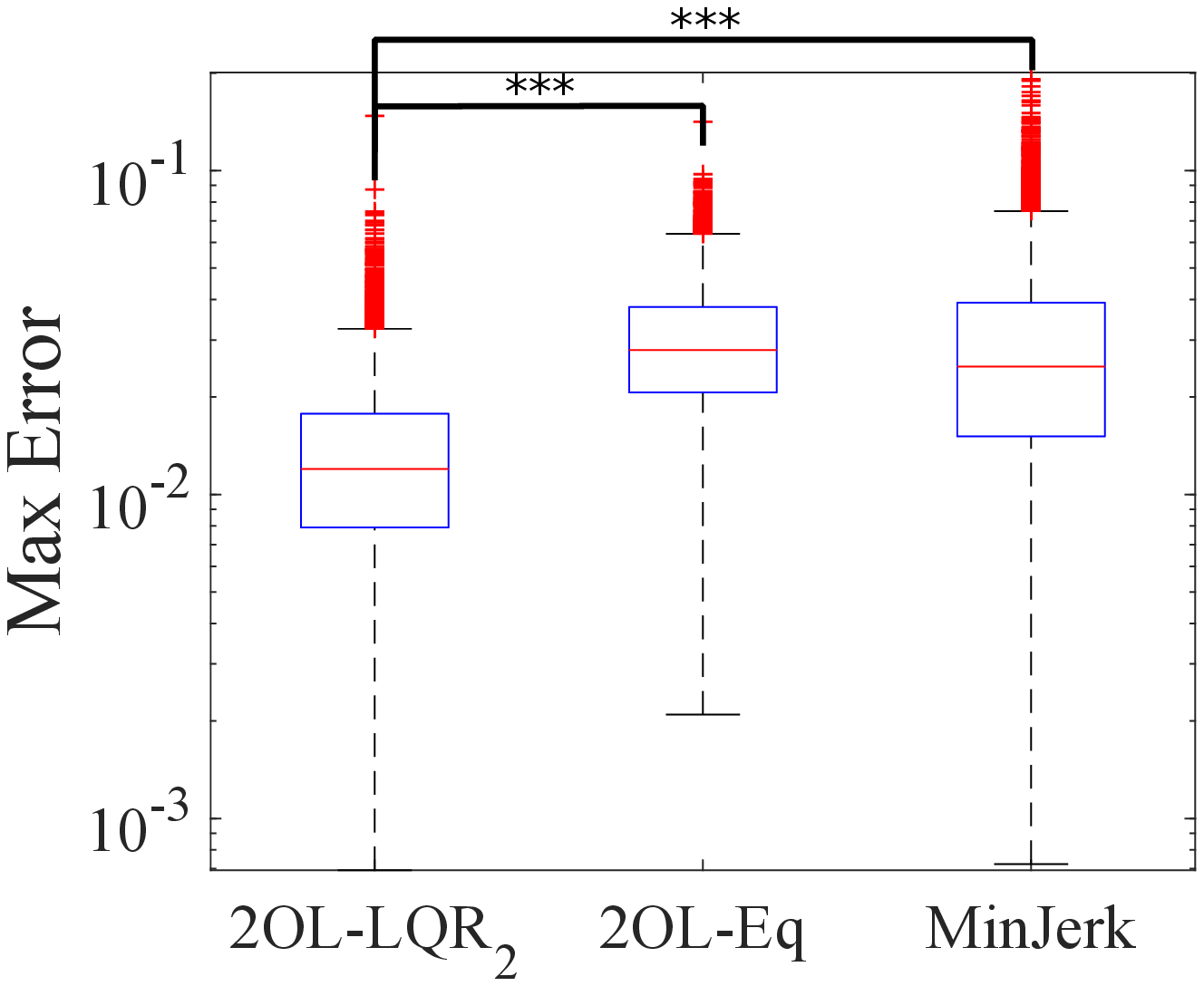}} 
	\caption{SSE and Maximum Error values of our model 2OL-LQR$_{2}$ compared to 2OL-Eq and MinJerk for the user trajectories of all participants and all tasks (logarithmic scale).}~\label{fig:boxplot-all-models}
\end{figure}

In the following, we provide a quantitative comparison across all 7702 trajectories.
The resulting SSE values of all three models are shown in Figure~\ref{fig:boxplot-all-models}(a), on a logarithmic scale.
In addition, we measure the \textit{Maximum Error} between model and user trajectories, i.e.,
\begin{align}
\max_{n=1,\dots,N}\vert p_{n}^{\Lambda}-p_{n}^{\text{USER}}\vert,
\end{align}
which is depicted in Figure~\ref{fig:boxplot-all-models}(b).
As can be seen from both Figures, our model 2OL-LQR$_2$ is able to capture human behavior substantially better in terms of SSE and in terms of Maximum Error than both the 2OL-Eq and MinJerk models.

Kolmogorov-Smirnov tests showed that the distributions of SSE for the three models do not fit the assumption of normality (all values $p<0.0001$). 
Thus, we carried out a Friedman Test (i.e., a non-parametric test equivalent to a repeated measures one-way ANOVA). 
The main factor included in the analysis was which model was used: 2OL-LQR$_{2}$, 2OL-Eq, or MinJerk. 
The significance level was set to $0.05$.
The test indicated that the SSE between the three models was significantly different ($\chi^{2}(2) = 8492.78$, $p<0.001$, $n=7702$). 

Additional Wilcoxon Signed Rank tests with Bonferroni corrections showed that the SSE was significantly lower in the 2OL-LQR$_{2}$ model when compared to the 2OL-Eq model ($Z=-74.87$, $p<0.001$), or to the MinJerk model\linebreak ($Z=-68.49$, $p<0.001$).
The findings are analogous for the maximum deviations of the simulated trajectories from the data (Friedman Test, $\chi^{2}(2) = 9106.12$, $p<0.001$, $n=7702$), with Wilcoxon Signed Rank tests ($p<0.001$) showing that 2OL-LQR$_{2}$ approximates user trajectories significantly better than both 2OL-Eq and MinJerk.
Summary statistics of both measures for all three models can be found in Table~\ref{tab:SSE-maxerror-properties}.
\begin{table} 
\center
\resizebox{\linewidth}{!}{
\begin{tabular}{|c|c|c|c|c|c|c|} 
\hline
\rule{0pt}{10pt}
\multirow{2}{*}{Model} & \multicolumn{3}{c|}{SSE} & \multicolumn{3}{c|}{Maximum Error} \\
\cline{2-7}
\rule{0pt}{10pt}
& Mean & SE & SD & Mean & SE & SD \\
\hline
\rule{0pt}{10pt}
2OL-LQR$_{2}$ & 0.03 & 0.001 & 0.10 & 0.014 & 0.0001 & 0.009  \\
\rule{0pt}{10pt}
2OL-Eq & 0.11 & 0.002 & 0.16 & 0.03 & 0.0001 & 0.013 \\
\rule{0pt}{10pt}
MinJerk & 0.21 & 0.006 & 0.56 & 0.035 & 0.0025 & 0.022 \\
\hline
\end{tabular}
}
\vspace*{0.05ex}
\caption{Mean value, standard error (SE), and standard deviation (SD) of the SSE and Maximum Error values of each model applied to the 7702 user trajectories.}\label{tab:SSE-maxerror-properties}
\end{table}

\subsection{Parameter Distribution of 2OL-LQR$_2$}
\begin{figure}[!t]
	\centering
	\subfloat[Parameter $k$]{
		\subfloat{\includegraphics[width=0.5\linewidth]{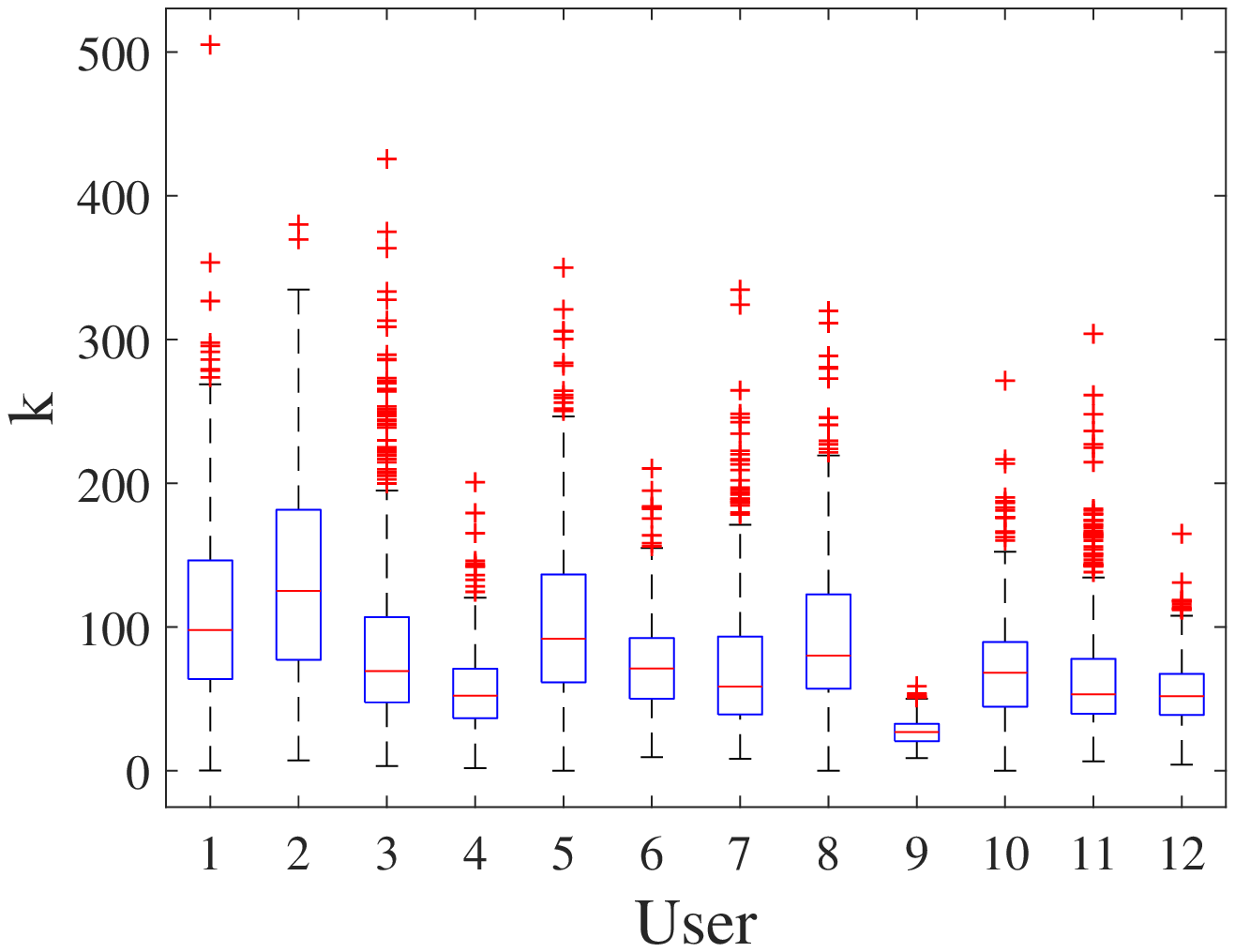}}
		\subfloat{\includegraphics[width=0.5\linewidth]{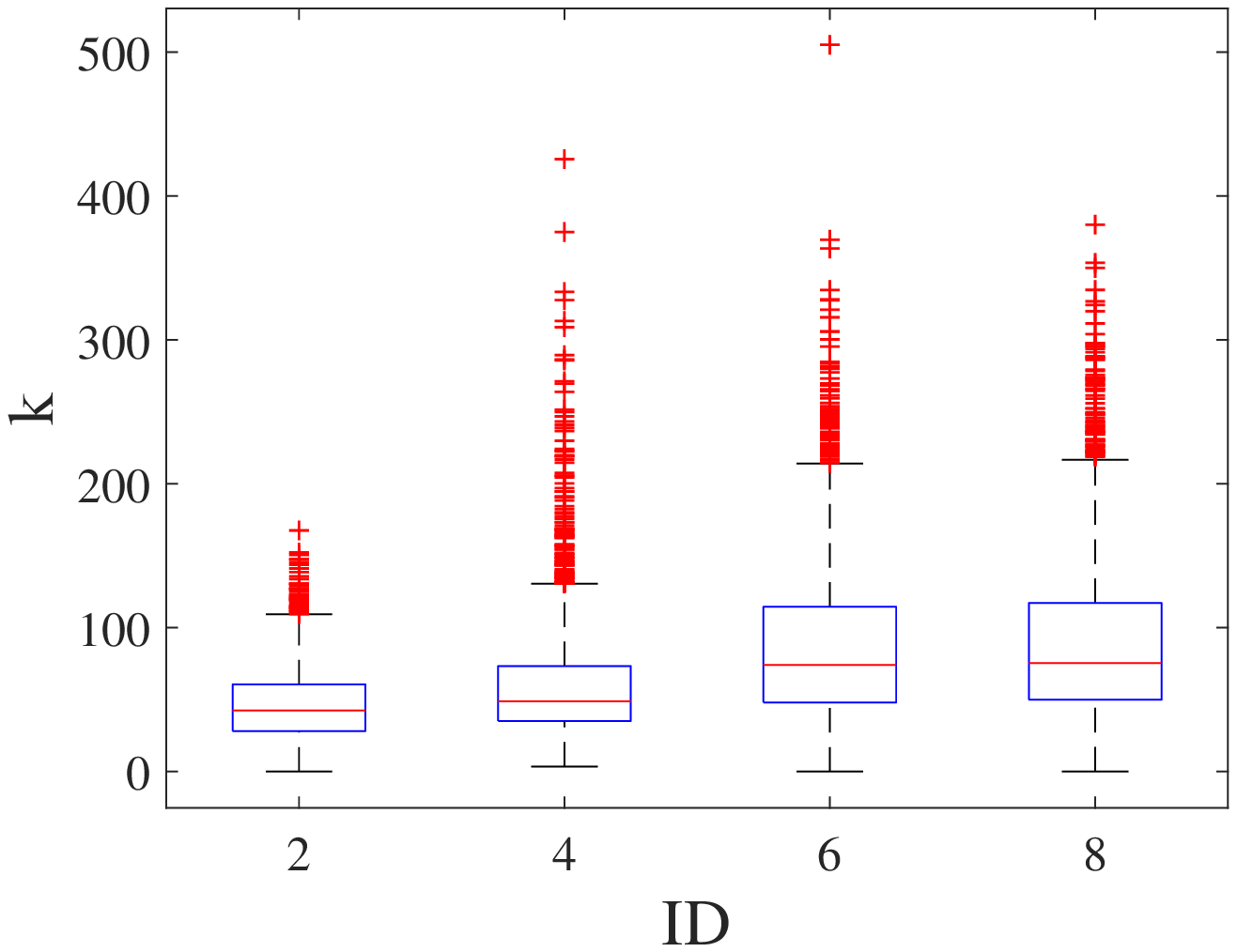}} 
		\addtocounter{subfigure}{-2}}
	\\
	\subfloat[Parameter $d$]{
		\subfloat{\includegraphics[width=0.5\linewidth]{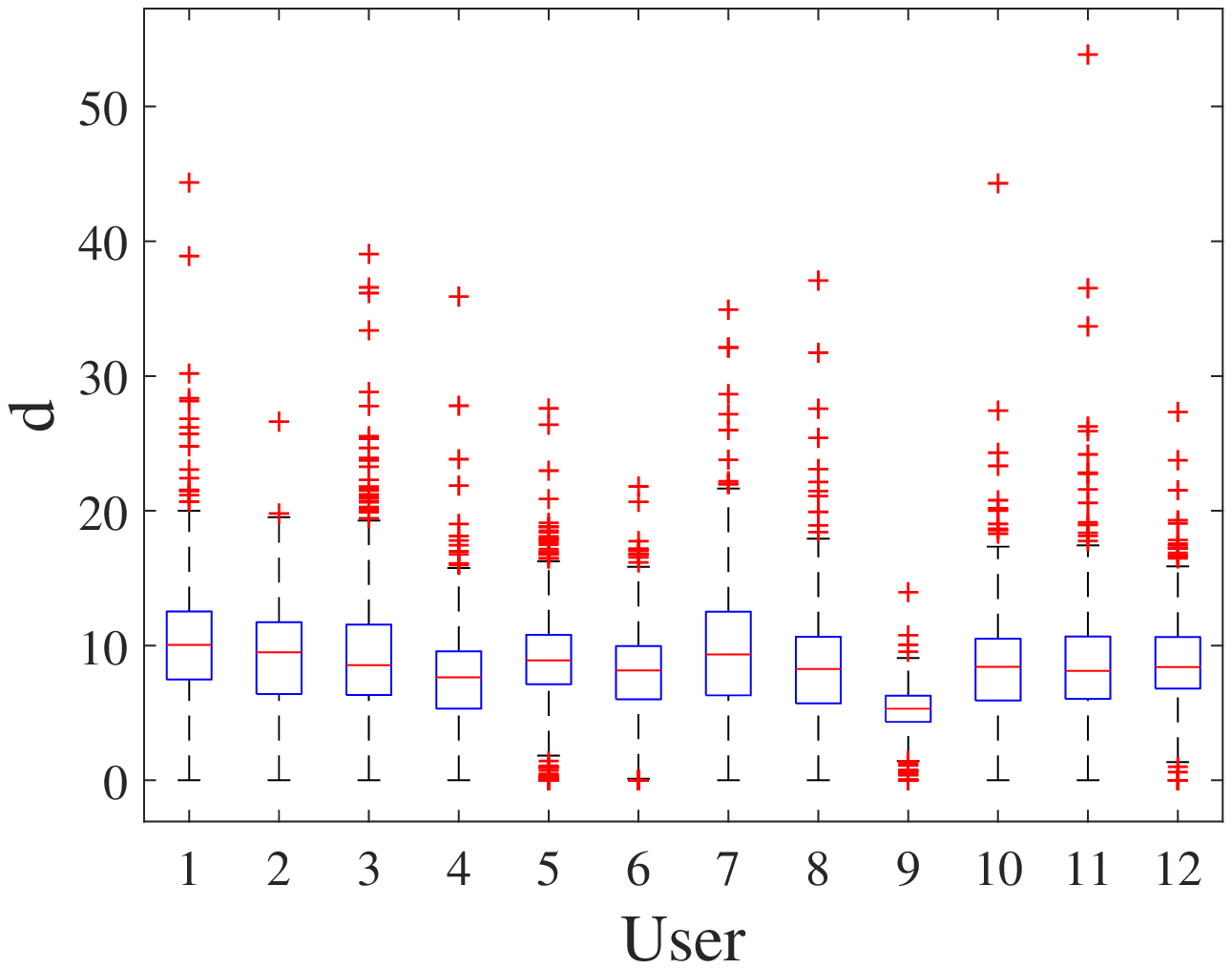}}
		\subfloat{\includegraphics[width=0.5\linewidth]{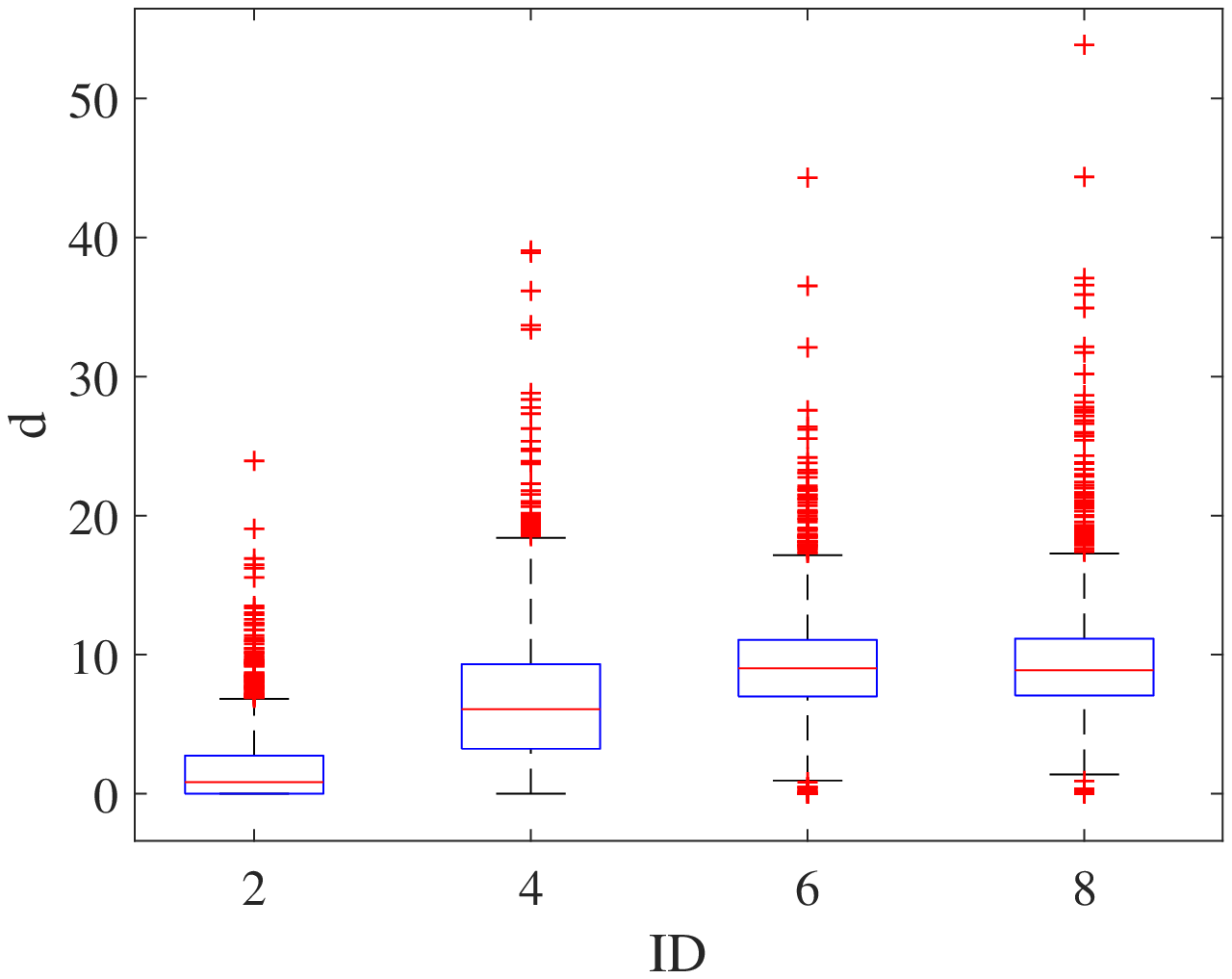}} 
		\addtocounter{subfigure}{-2}}
	\\
	\subfloat[Parameter $r$ (logarithmic scale)]{
		\subfloat{\includegraphics[width=0.5\linewidth]{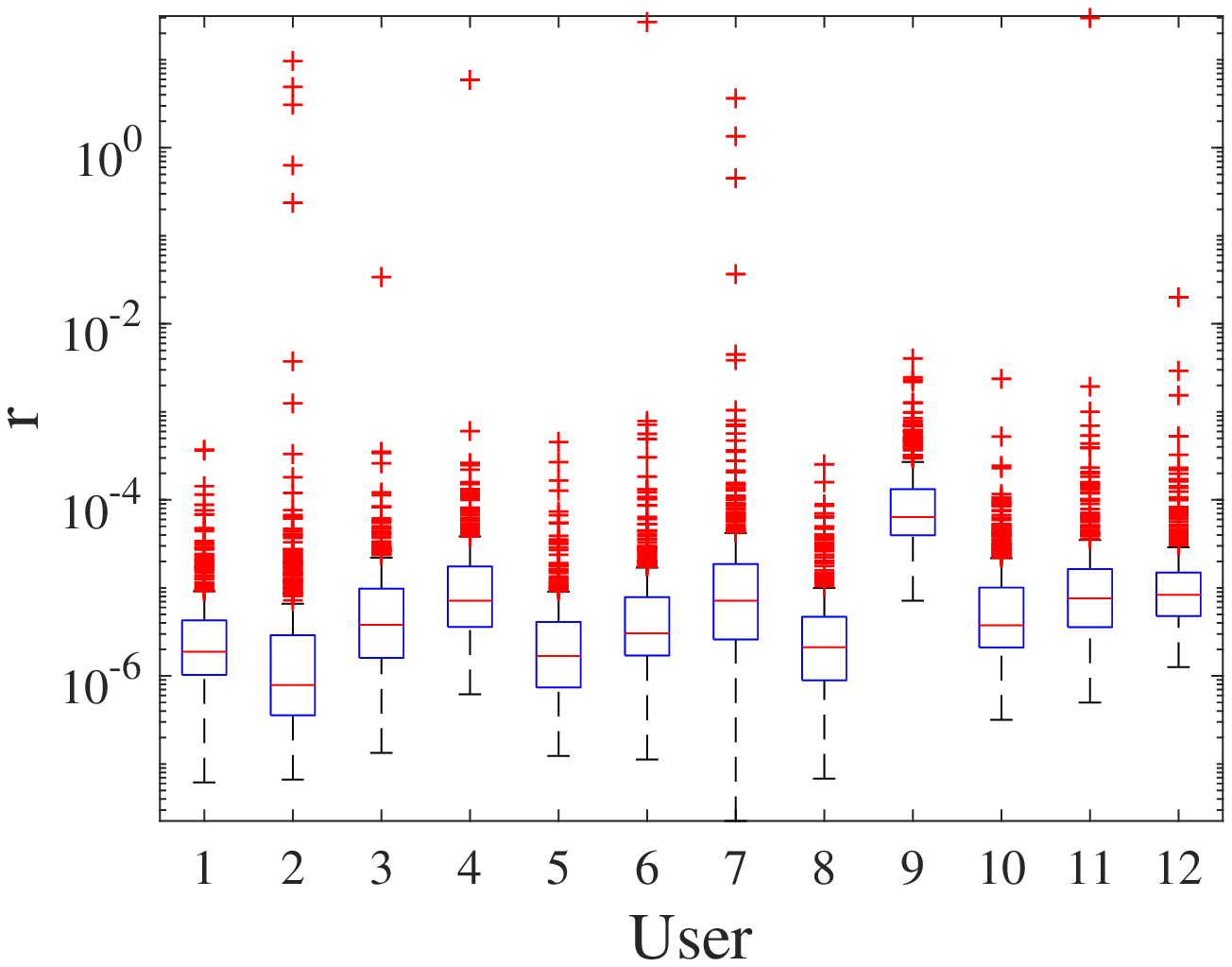}}
		\subfloat{\includegraphics[width=0.5\linewidth]{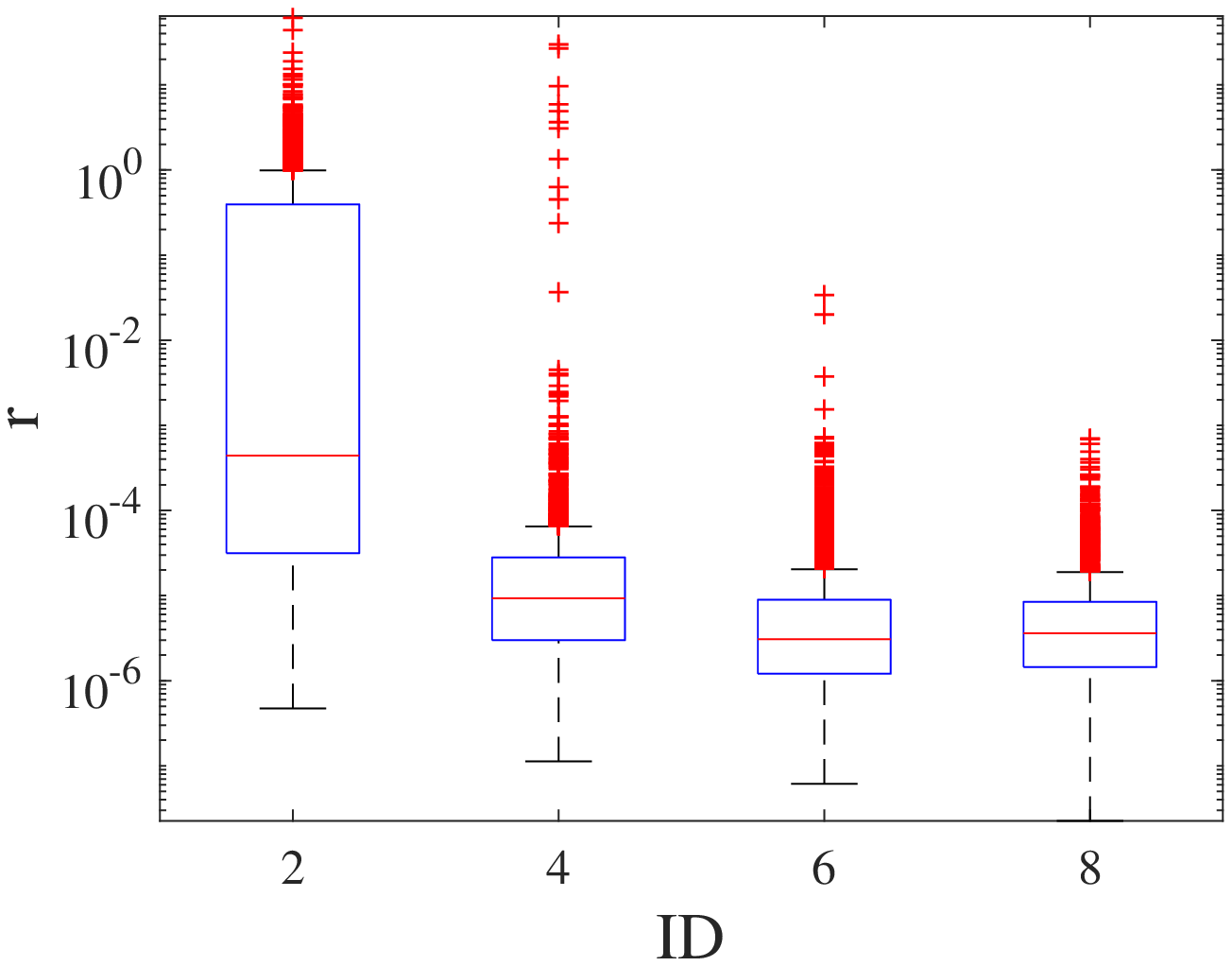}}
		\addtocounter{subfigure}{-2}}
	\caption{Parameters of our model 2OL-LQR$_{2}$, optimized for all considered trajectories of all participants and all tasks, grouped by participants (left, only ID 4, 6, 8 tasks) and by ID (right). For reasons of clarity, both plots for parameter $d$ do not include the five biggest outliers ranging between 58 and 181.}~\label{fig:boxplot-var2}
\end{figure}
Figures~\ref{fig:boxplot-var2}(a)-(c) (left) show the ranges of the three 2OL-LQR$_2$ parameters $k$, $d$, and $r$, optimized for the user trajectories of all tasks with $\text{ID}>2$, grouped by participants.\footnote{
The parameters for ID~2 tasks differ from those of $\textnormal{ID}>2$ tasks. 
Due to limited space, we focus on the latter in these plots.
For the sake of completeness, the figures including ID~2 tasks can be found in the supplementary material.}
As can be seen, different participants are characterized by differing parameter sets.
For example, participant 2 is characterized by a high spring stiffness $k$, an above-average damping~$d$, and a very low jerk weight~$r$.
In contrast, participant 9 is characterized by a very low spring stiffness~$k$, a very low damping~$d$, and a very high jerk weight~$r$.
Since in our case higher jerk penalization enforces less rapid changes in control, from the jerk weight~$r$ it can be inferred how much \emph{effort} the user is willing to put into the task: a higher $r$ can be interpreted as less effort.

Figures~\ref{fig:boxplot-var2}(a)-(c) (right) illustrate the ranges of the parameters $k$, $d$, and $r$, optimized for the user trajectories of all participants, grouped by ID of the task. 
All three parameters show characteristic variations by ID.
The spring stiffness $k$ increases noticeably from ID~4 to ID~6.
The damping parameter $d$ is considerably lower for ID~2 tasks.
This confirms the observation that participants show oscillatory behavior in tasks with low IDs, as reported before in \cite{Gui93,BooReiFer04,mueller17}.
These oscillations also play a role in the large variance of $r$ for ID~2.
For the other IDs, $r$ declines only slightly with ID, i.e., the effort is almost independent of the task difficulty.

The impact of the parameters on model behavior is however not straightforward, because a change in one of the parameters does not only influence the movement directly, but also results in a different optimal control sequence, which likewise affects the solution trajectory.

\subsection{Modeling Individual Movements Including Reaction Time}
\begin{figure}
	\centering
	\subfloat[Position Time Series]{\includegraphics[width=0.5\linewidth]{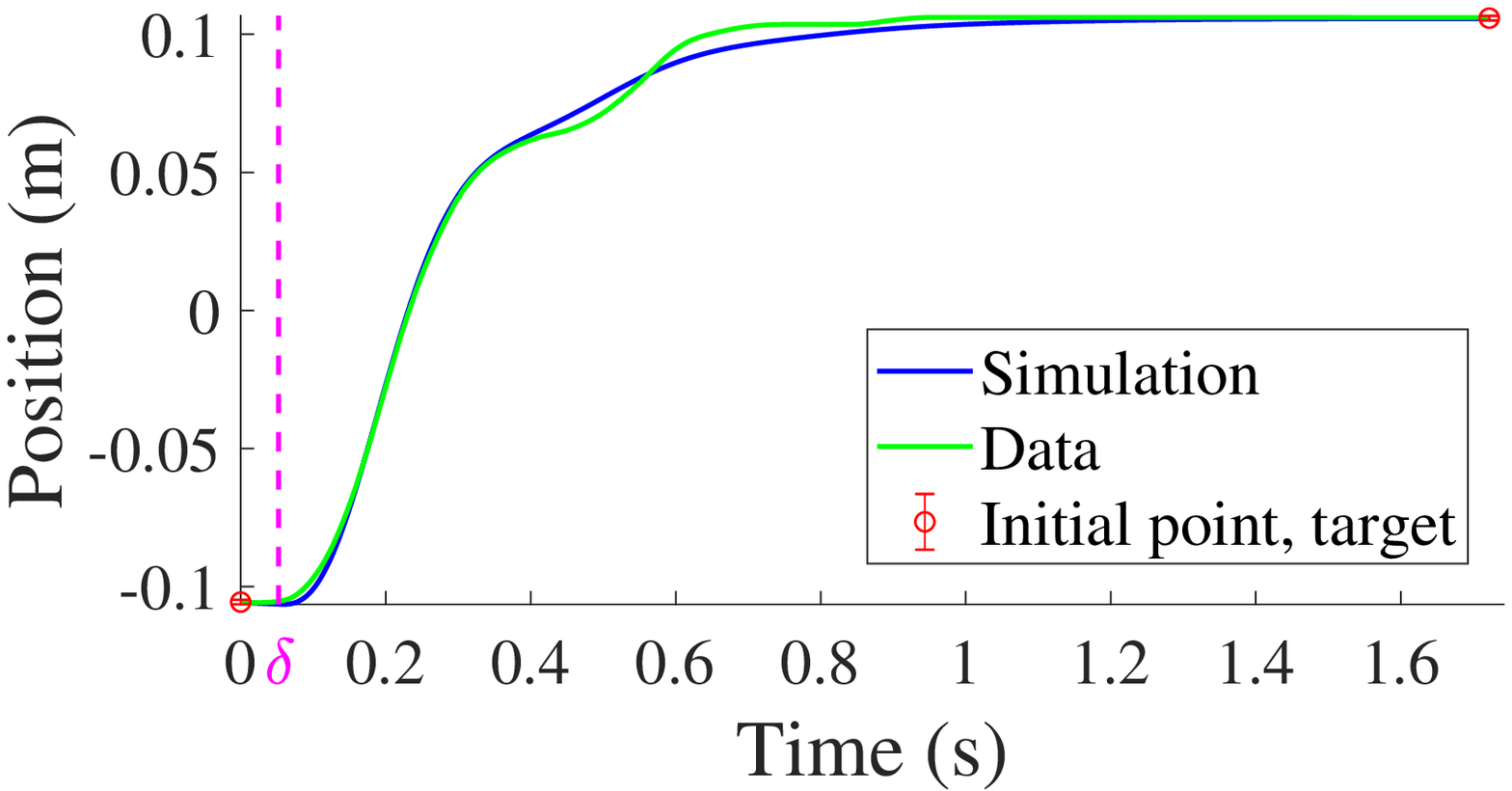}}
	\subfloat[Velocity Time Series]{\includegraphics[width=0.5\linewidth]{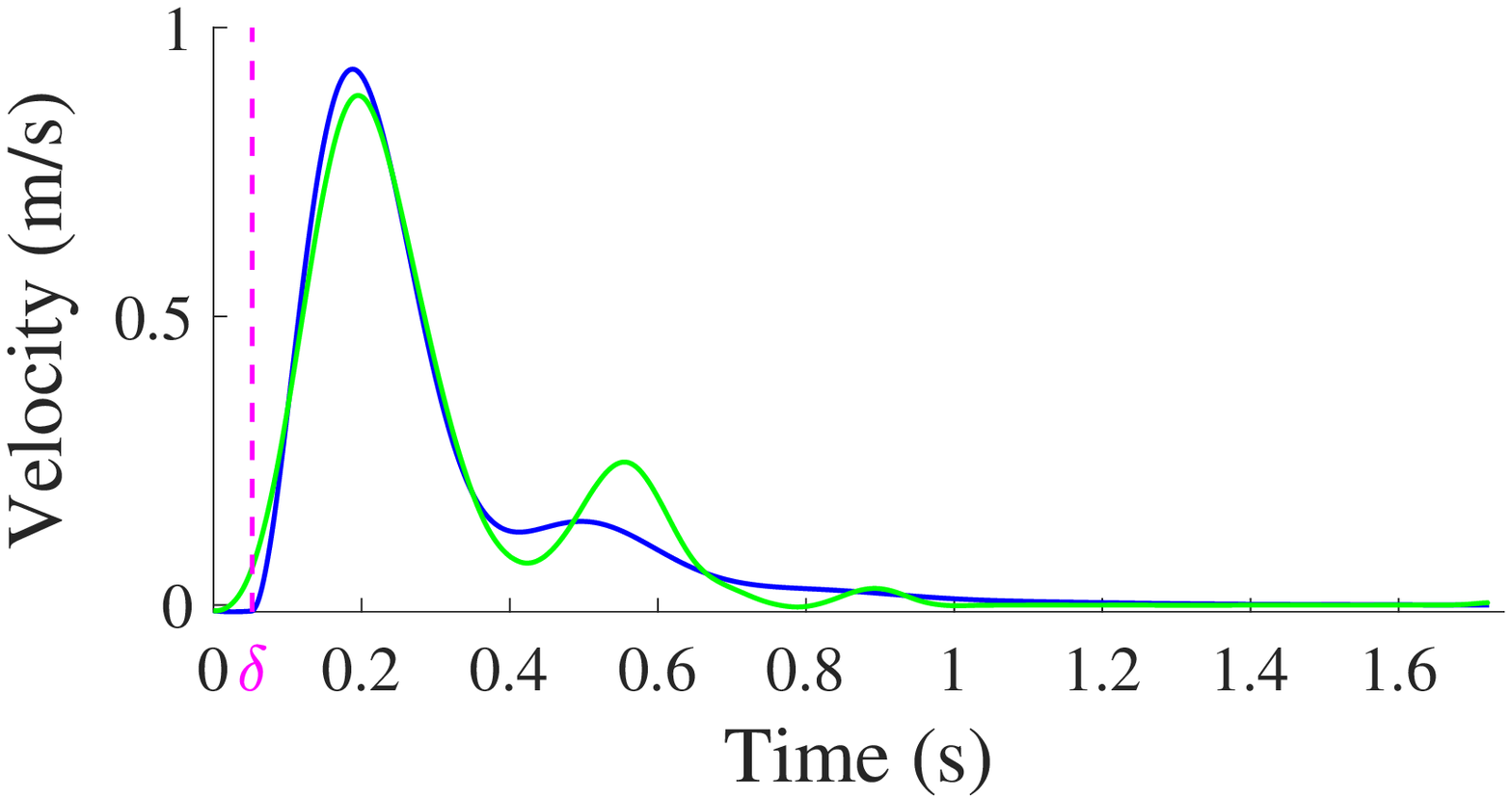}}
	\\
	\subfloat[Acceleration Time Series]{\includegraphics[width=0.5\linewidth]{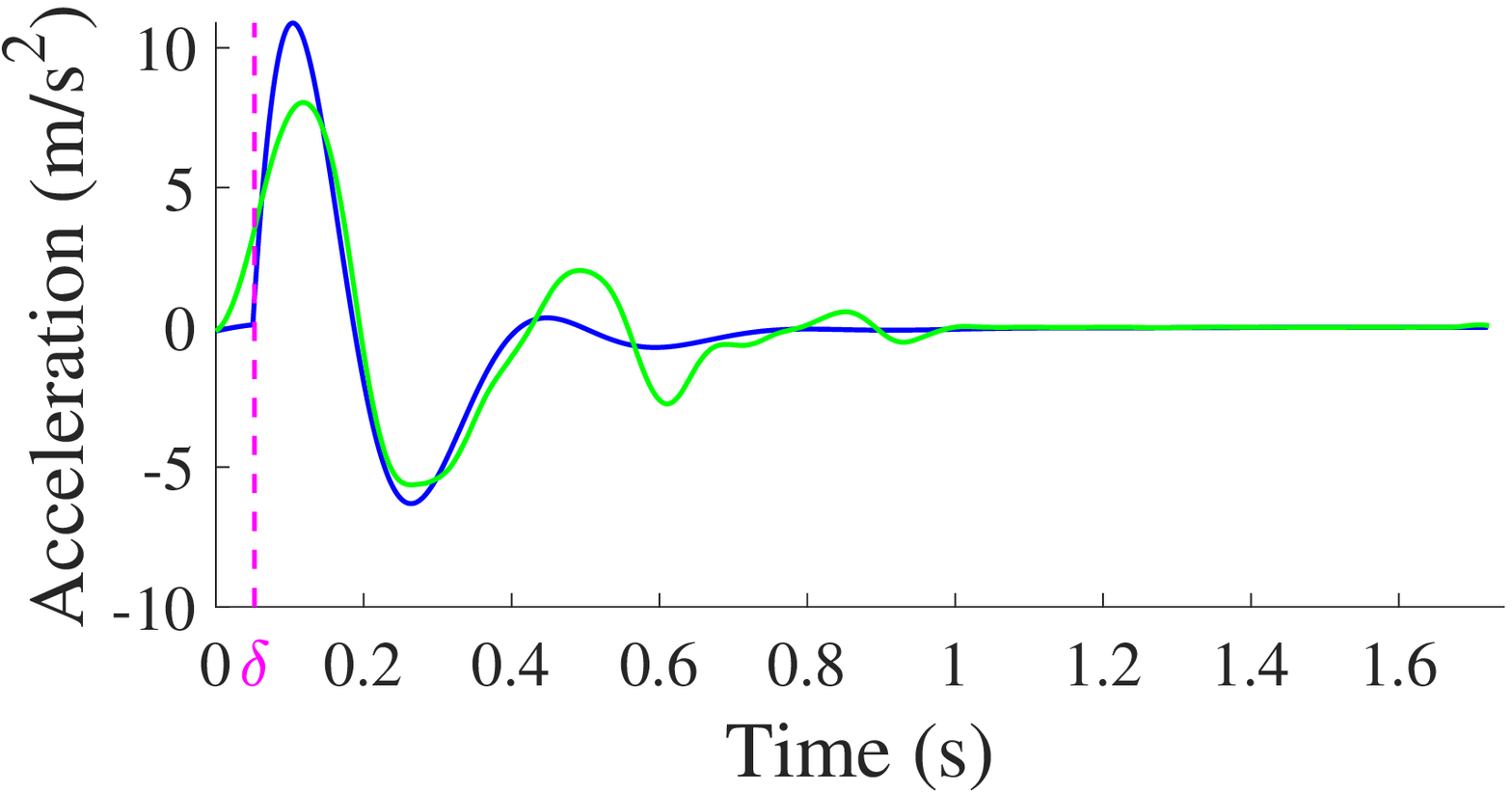}}
	\subfloat[Control Time Series]{\includegraphics[width=0.5\linewidth]{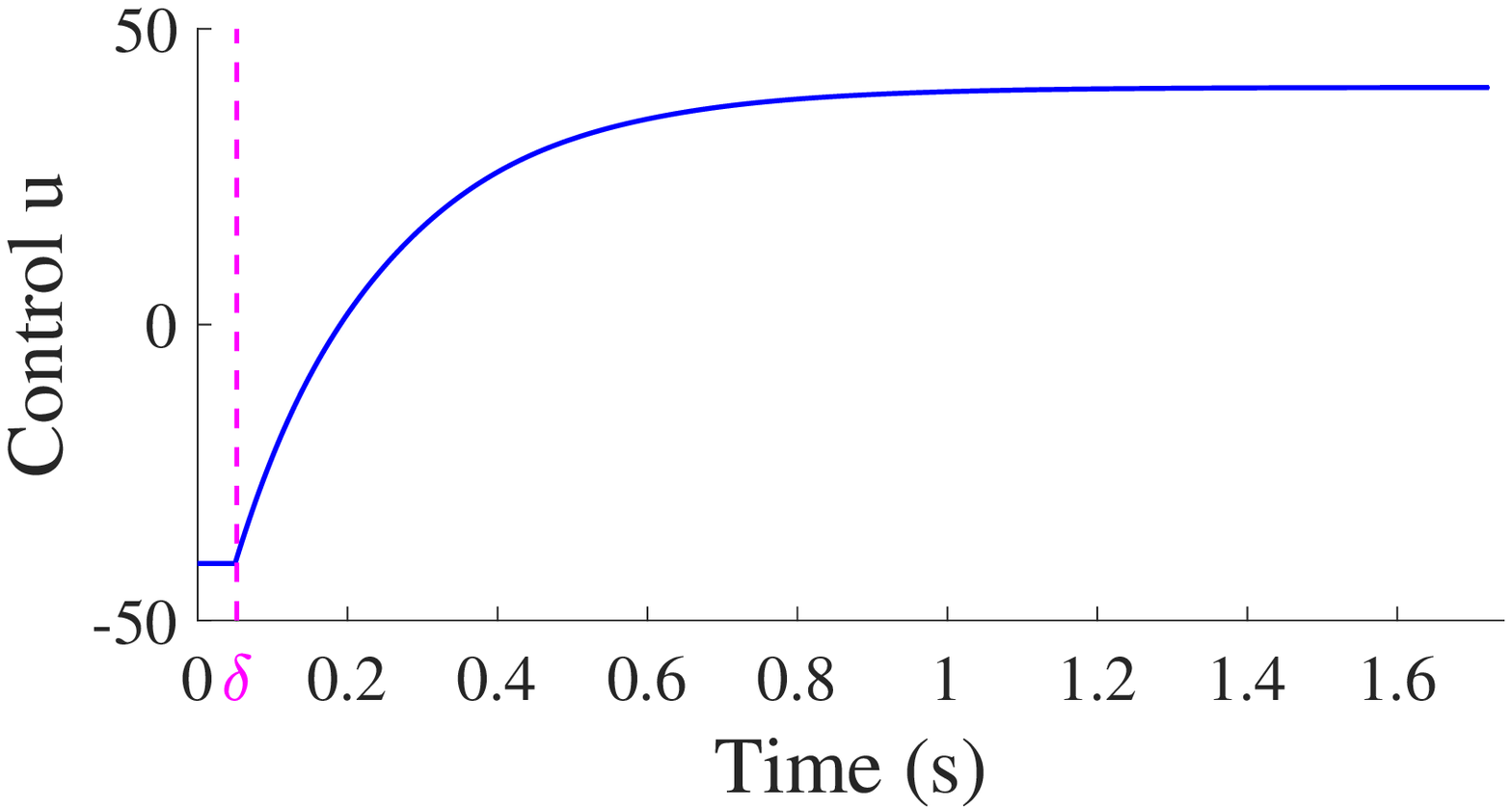}}
	\caption{Our third iteration model 2OL-LQR$_{3}$ allows to model individual movements by including reaction time.}~\label{fig:var3-rt}
\end{figure}
Our model 2OL-LQR$_2$ does not take reaction time into account.
However, this is possible with our third iteration, 2OL-LQR$_{3}$.
Only in this section, we thus explicitly do \textit{not} drop any frames at the beginning of the trials.
Results for the same representative trial as before are shown in Figure~\ref{fig:var3-rt}.
Clearly, there is no change in control and thus in acceleration before time $\delta$, which can loosely be interpreted as a reaction time.
Looking closely at the initiation of the acceleration, we observe that our model initiates the movement later than the user but with a higher acceleration.
The reason is that the optimizer treats $\delta$ as a free parameter to minimize the SSE of the entire position time series. 
Thus, while movements including reaction time can be approximated by 2OL-LQR$_3$ quite well, the parameter~$\delta$ itself does not necessarily resemble the true reaction time.

\section{Discussion and Future Work}
In this paper we have explored a simple OFC model for mouse pointer movements.
We assumed \emph{optimal closed-loop behavior} with respect to a \emph{quadratic cost function} (penalizing jerk and distance) and subject to \emph{linear system dynamics} with {\em no delay} and \emph{no noise}.
These simplifications lead to a number of limitations of our model. 

First, all models that we compared do not model corrective submovements well.
Although our models can recreate corrective submovements (e.g., in Figure~\ref{fig:var3-rt}), they are smaller in amplitude than those of the users.
Future research should put more emphasis on replicating these submovements in more detail by extending the model. 

Second, due to its deterministic nature, our model cannot replicate the variability of human movements.
It produces a typical movement of a specific user, but it produces the same movement every time.
In future work we plan to explore stochastic models to better capture human variability. 

Third, we note that although our cost function~\eqref{eq:Cost Functional3} of our main model, 2OL-LQR$_2$, incentivizes a short(er) movement time due to summed distance costs, it does not explicitly model minimizing the total movement time. 
If the latter is desired (e.g., as part of the experimental design), then in future work the model can be extended by modifying the cost function using the Cost of Time theory.

Despite these limitations, our 2OL-LQR$_2$ model matches our data well, and significantly better than 2OL-Eq or MinJerk.
We achieve this with only three parameters, which have 
an easily understandable interpretation as spring stiffness~$k$, damping~$d$, and effort, related to~$r$.
We only need these parameters, the target position, and initial conditions.
In contrast to MinJerk, our model does not need to know the point in time and space where the surge movement ends.
Most importantly, our model does not require knowledge about the exact time when the target is reached.
Compared to 2OL-Eq, our model yields a more bell-shaped velocity time series and a more N-shaped acceleration time series, without implausibly high acceleration at the start of the movement.
In addition, our model explains how users differ from each other in properties (stiffness, damping) and effort.

The biggest strength is that the OFC perspective makes our model very flexible and easily extensible.
In particular, it can readily be extended to other instructions, such as emphasizing speed vs.\ comfort.
It can also be extended to different tasks, such as 2D or 3D pointing, 6 DoF docking tasks, etc. 

It is important to highlight that our model is a pure end-effector model of the movement of the mouse pointer. 
We do not explicitly model biomechanics, sensor characteristics, or transfer functions in the operating system.
Incorporating these is possible, albeit yielding nonlinear system dynamics, and therefore making the model more complex. 
Our simple model already works quite well for modeling mouse pointer movements.
This reinforces our argument that OFC is a promising theory to better understand movement, such as movement of the mouse pointer, during interaction and is thus a valuable addition to the HCI community.

\section{Conclusion}
In this paper, we have modeled mouse pointer movements from an optimal control perspective. 
More precisely, we have investigated the Linear-Quadratic Regulator with various objective functions.
We found that our model 2OL-LQR$_2$ fits our data significantly better than either 2OL-Eq~\cite{mueller17} or MinJerk~\cite{Flash85}. 
We require a number of simplifying assumptions (linear dynamics, quadratic costs).
Despite these, mouse pointer movements of real users can be explained well.
Moreover, this is achieved with only three, intuitively interpretable, parameters, which allow to characterize users by properties (stiffness, damping) and effort.
In conclusion, we believe that the optimal feedback control perspective is a strong, flexible, and very promising direction for HCI, which should be further explored in the future.

\balance{}

\bibliographystyle{SIGCHI-Reference-Format}
\bibliography{bibliography-long}

\appendix

\section{2OL-LQR equations}

The 2OL-LQR model can be described as the time-discrete linear-quadratic optimal control problem with finite horizon $N\in\N$
\begin{subequations}\label{eq:2OL-LQR}
	\begin{small}
	\begin{align}\label{eq:2OL-LQR-objective}
	\begin{gathered}
	\textsl{Minimize} \quad J_{N}(x,u)= \sum_{n=1}^{N}x_{n}^{\top}Q_{n}x_{n} + \sum_{n=1}^{N-1} (u_{n}-u_{n-1})^{\top}R_{n}(u_{n}-u_{n-1}) \\
	\textsl{with respect to } u= (u_{n})_{n\in\{1,\dots,N-1\}}\subset \R \textsl{ given } \bar{x}_{1}\in\R^{3}, \, \bar{u}_{0}\in\R
	\end{gathered} 
	\end{align}
	\end{small}
	where $x= (x_{n})_{n\in\{1,\dots,N\}} \subset \R^{3}$ with $x_{n}= \left( p_{n},v_{n},T \right)^{\top}$ satisfies
	\begin{align}\label{eq:discrete-control_3}
	\begin{gathered}
	x_{n+1}=A x_{n} + B u_{n}, \quad n\in\{1,\dots,N-1\}, \\
	x_{1}=\bar{x}_{1},
	\end{gathered}
	\end{align}
	with sampling time $h>0$ and system dynamics matrices
	\begin{align}
	A= \begin{pmatrix}
	1 & h & 0 \\
	-hk & 1-hd & 0 \\
	0 & 0 & 1
	\end{pmatrix}, \quad B= \begin{pmatrix}
	0 \\
	h \\
	0
	\end{pmatrix}
	\end{align}
	\end{subequations}
	based on the (approximated) second-order lag.
	\\
	The state cost matrices are defined by
	\begin{align}\label{eq:2OL-LQR-cost-matrices_2}
	Q_{n}= 
	\begin{pmatrix}
	1 & 0 & -1 \\
	0 & 0 & 0 \\
	-1 & 0 & 1
	\end{pmatrix}\in\R^{3\times 3}, \quad n\in\{1,\dots,N\},
	\end{align}
	which implies
	\begin{align}
	x_{n}^{\top}Q_{n}x_{n}=(T-p_{n})^{2} = D_{n}^{2},
	\end{align}
	i.e., the distance $D_{n}=\vert T - p_{n} \vert$ between mouse and target position is quadratically penalized at every time step $n\in\{1,\dots,N\}$.
	In our case of one-dimensional pointing tasks, the control cost matrices are scalar and given by
	\begin{align}\label{eq:2OL-LQR-jerk-weights_2}
	R_{n}=\frac{r}{h^{2}} \in\R, \quad r>0, \quad n\in\{1,\dots,N-1\},
	\end{align}
	which yields \begin{align}
	(u_{n}-u_{n-1})^{\top}R_{n}(u_{n}-u_{n-1})=r_{n}\left(\frac{u_{n}-u_{n-1}}{h}\right)^{2},
	\end{align}
	i.e., the squares of the ``jerk'' terms $j_{n}=\frac{u_{n}-u_{n-1}}{h}$ are penalized with some jerk weight $r$ at every time step $n\in\{1,\dots,N-1\}$.
	\\
	Because of the penalization of the \textit{differences} in control, each control value $u_{n}^{*}$ of the optimal control sequence $u^{*}$ minimizing $J_{N}(x,u)$ given some initial state $\bar{x}_{1}$ and some initial control $\bar{u}_{0}$ explicitly depends on the preceding control value $u_{n-1}^{*}$.
	For this reason, we need to introduce \textbf{information vectors}
	\begin{align}\label{eq:information-vectors}
	\mathcal{I}_{n}= \begin{pmatrix} x_{n} \\ u_{n-1} \end{pmatrix}\in\R^{4}, \quad n\in\{1,\dots,N\}.
	\end{align}
	Furthermore, we expand the system matrices $A$ and $Q_{n}$ by an additional zero row and column and add an additional one to the control matrix $B$ in order to propagate the previous control $u_{n-1}$:	\begin{align}\label{eq:discrete-control-matrices_information-vectors}
	\begin{gathered}
	\mathcal{A}= \begin{pmatrix}
	A & 0 \\
	0 & 0 
	\end{pmatrix} = \begin{pmatrix}
	1 & h & 0 & 0\\
	-hk & 1-hd & 0 & 0 \\
	0 & 0 & 1 & 0 \\
	0 & 0 & 0 & 0
	\end{pmatrix}\in\R^{4\times 4}, \\
	\mathcal{B}= \begin{pmatrix}
	B \\
	1
	\end{pmatrix} = \begin{pmatrix}
	0 \\
	h \\
	0 \\
	1
	\end{pmatrix}\in\R^{4\times 1}, \\
	\mathcal{Q}_{n} = \begin{pmatrix}
	Q_{n} & 0 \\
	0 & 0
	\end{pmatrix} = \begin{pmatrix}
	1 & 0 & -1 & 0 \\
	0 & 0 & 0 & 0 \\
	-1 & 0 & 1 & 0 \\
	0 & 0 & 0 & 0
	\end{pmatrix}\in\R^{4\times 4},
	\end{gathered} \nonumber \\
	n\in\{1,\dots,N\}.
	\end{align}
	\\
	Using this notion, \eqref{eq:2OL-LQR} is equivalent to the following optimal control problem:
	\begin{subequations}\label{eq:2OL-LQR_informationvector}
	\begin{small}
	\begin{align}\label{eq:2OL-LQR-objective_information-vectors}
	\begin{gathered}
	\textsl{Minimize} \quad \mathcal{J}_{N}(\mathcal{I},u)= \sum_{n=1}^{N}\mathcal{I}_{n}^{\top}\mathcal{Q}_{n}\mathcal{I}_{n} + \sum_{n=1}^{N-1} (u_{n}-u_{n-1})^{\top}R_{n}(u_{n}-u_{n-1}) \\
	\textsl{with respect to } u= (u_{n})_{n\in\{1,\dots,N-1\}}\subset \R \textsl{ given } \bar{x}_{1}\in\R^{3}, \, \bar{u}_{0}\in\R
	\end{gathered} 
	\end{align}
	\end{small}
	where $\mathcal{I}= (\mathcal{I}_{n})_{n\in\{1,\dots,N\}} \subset \R^{4}$ with $\mathcal{I}_{n}= \left( x_{n}, u_{n-1} \right)^{\top}$ satisfies
	\begin{align}\label{eq:discrete-control_information-vectors}
	\begin{gathered}
	\mathcal{I}_{n+1}=\mathcal{A}\mathcal{I}_{n}+\mathcal{B}u_{n}, \quad n\in\{1,\dots,N-1\}, \\
	\mathcal{I}_{1}=\bar{\mathcal{I}}_{1} = \begin{pmatrix} \bar{x}_{1} \\ \bar{u}_{0} \end{pmatrix},
	\end{gathered}
	\end{align}
	\end{subequations}
	with sampling time $h>0$ and where $u_{0}=\bar{u}_{0}$ applies.
	\\
	Moreover, we define 
	\begin{align}\label{eq:information-vector-extractors}
	\mathrm{I}_{x}= \begin{pmatrix}
	1 & 0 & 0 & 0 \\
	0 & 1 & 0 & 0 \\
	0 & 0 & 1 & 0
	\end{pmatrix}\in\R^{3\times 4}, \quad
	\mathrm{I}_{u}= \begin{pmatrix}
	0 & 0 & 0 & 1
	\end{pmatrix}\in\R^{1\times 4},
	\end{align} 
	which implies
	\begin{align}\label{eq:information-vector-extraction}
	\mathrm{I}_{x}\mathcal{I}_{n}=x_{n}\in\R^{3}, \quad \mathrm{I}_{u}\mathcal{I}_{n}=u_{n-1}\in\R, \quad n\in\{1,\dots,N\},
	\end{align}
	i.e., $\mathrm{I}_{x}$ respective $\mathrm{I}_{u}$ are the matrices that extract the state $x_{n}$ respective the control $u_{n-1}$ from the information vector $\mathcal{I}_{n}$ for any $n\in\{1,\dots,N\}$.
	\bigbreak
	It can be shown that the unique solution $u^{*}=(u_{n}^{*})_{n\in\{1,\dots,N\}}$ to the optimization problem \eqref{eq:2OL-LQR_informationvector} (and thus to the original optimization problem \eqref{eq:2OL-LQR} as well) is given by
	\begin{align}\label{eq:2OL-LQR-feedback}
	\begin{gathered}
	u_{n}^{*}= -K_{n} \mathcal{I}_{n}^{*}, \quad n\in\{1,\dots,N-1\}, \\
	K_{n}=(R_{n}+\mathcal{B}^{\top}\mathcal{S}_{n+1}\mathcal{B})^{-1}(\mathcal{B}^{\top}\mathcal{S}_{n+1}\mathcal{A}-R_{n}\mathrm{I}_{u}),
	\end{gathered} \nonumber \\
	n\in\{1,\dots,N-1\},
	\end{align}
	where the symmetric matrices $\mathcal{S}_{n}\in\R^{4\times 4}$ can be determined by solving the \textbf{Modified Discrete Riccati Equations}
	\begin{subequations}\label{eq:modified-discrete-riccati}
		\begin{align}
		\begin{gathered}
		\mathcal{S}_{n}= \mathcal{Q}_{n} + \mathrm{I}_{u}^{\top}R_{n}\mathrm{I}_{u} + \mathcal{A}^{\top}\mathcal{S}_{n+1}\mathcal{A} - \\
		- (\mathcal{A}^{\top}\mathcal{S}_{n+1}\mathcal{B} - \mathrm{I}_{u}^{\top}R_{n})(R_{n} + \mathcal{B}^{\top}\mathcal{S}_{n+1}\mathcal{B})^{-1}(\mathcal{B}^{\top}\mathcal{S}_{n+1}\mathcal{A} - R_{n}\mathrm{I}_{u})
		\end{gathered}
		\end{align}
		for $n\in\{1,\dots,N-1\}$ backwards in time with initial value 
		\begin{align}
		\mathcal{S}_{N}=\mathcal{Q}_{N}.
		\end{align}
	\end{subequations}
\end{document}